\documentclass[12pt]{article}
\usepackage[utf8]{inputenc}
\usepackage[style=authoryear]{biblatex}
\usepackage{csquotes}
\usepackage[utf8]{inputenc}
\usepackage[T1]{fontenc}
\usepackage{float}
\usepackage{pbox}
\usepackage{wrapfig}
\usepackage{xcolor}
\usepackage[title]{appendix}
\usepackage{amsmath}
\usepackage{amsfonts}
\usepackage{mathtools}
\usepackage{bm}
\usepackage{url}
\usepackage{hyperref}
\usepackage[margin=0.75in]{geometry}
\usepackage[affil-it]{authblk}

\newcommand{\E}{\mathbb{E}}
\newcommand{\V}{\text{Var}}
\newcommand{\given}{\:|\:}
\newcommand{\comment}[1]{}

\title{Inferring Density-Dependent Population Dynamics Mechanisms through Rate Disambiguation for Logistic Birth-Death Processes}

\author[1,*]{Linh Huynh}
\author[2,3]{Jacob G. Scott}
\author[1,4,5]{Peter J. Thomas}
\affil[1]{Department of Mathematics, Applied Mathematics, and Statistics, Case Western Reserve University, Cleveland, OH, 44106 USA}
\affil[2]{Department of Translational Hematology and Oncology Research, Cleveland Clinic, Cleveland, OH, 44106 USA}
\affil[3]{Department of Systems Biology and Bioinformatics, School of Medicine, Case Western Reserve University, Cleveland, OH, 44106 USA}
\affil[4]{Department of Biology, Case Western Reserve University, Cleveland, OH, 44106 USA}
\affil[5]{Department of Computer and Data Science, Case Western Reserve University, Cleveland, OH, 44106 USA}
\affil[*]{Corresponding author: lxh390@case.edu}

\begin{document}
\flushbottom
\maketitle
\begin{abstract}
Density dependence is important in the ecology and evolution of microbial and cancer cells.
Typically, we can only measure net growth rates, but the underlying density-dependent mechanisms that give rise to the observed dynamics can manifest in birth processes, death processes, or both.
Therefore, we utilize the mean and variance of cell number fluctuations to separately identify birth and death rates from time series that follow stochastic birth-death processes with logistic growth.
Our method provides a novel perspective on stochastic parameter identifiability, which we validate by analyzing the accuracy in terms of the discretization bin size.
We apply our method to the
scenario where a homogeneous cell population goes through three stages: (1)
grows naturally to its carrying capacity, (2) is treated with a drug that
reduces its carrying capacity, and (3) overcomes the drug effect to restore
its original carrying capacity. 
In each stage, we disambiguate whether it
happens through the birth process, death process, or some combination of the two, which contributes to understanding drug resistance mechanisms.
In the case of limited data sets, we provide an alternative method based on maximum likelihood and solve a constrained nonlinear optimization problem to identify the most likely density dependence parameter for a given cell number time series. Our methods can be applied to other biological systems at different scales to disambiguate density-dependent mechanisms underlying the same net growth rate. 
\end{abstract}

\paragraph{Acknowledgements}
This work was made possible in part by NSF grant DMS-2052109, by research support from the Oberlin College Department of Mathematics, the National Institutes of Health (5R37CA244613-02), and the American Cancer Society Research Scholar Grant (RSG-20-096-01).
The authors thank Dr.~Vishhvaan Gopalakrishnan and Mina Dinh for sharing preliminary experimental data from the  EVE system (EVolutionary biorEactor), and Dr.~Kyle Card for discussing the problem of calibrating optical density measurements.
\paragraph{Statements and Declarations} The authors have declared no competing interest. 
Code is available at \url{https://github.com/lhuynhm/birthdeathdisambiguation}.
\newpage
\section{Introduction}\label{sect:introduction}
Density dependence, a phenomenon in which a population's \textit{per capita} growth rate changes with population density \parencite{Hixon:2009}, plays an important role in the ecology and evolution of microbial populations or tumors, especially under drug treatments. 
For example, Karslake et al. 2016 \parencite{karslake2016population} shows experimentally that changes in \textit{E.coli.} cell density can either increase or decrease the efficacy of antibiotics. 
Existing work such as \parencite{kaznatcheev2019fibroblasts}, \parencite{paczkowski2021reciprocal}, \parencite{emond2021ecological}, \parencite{susswein2022borrowing}, and \parencite{farrokhian2022measuring} shows that interactions between drug sensitive and resistant cancerous cells can shape the population's evolution of drug resistance. 
To analyze the role of density dependence, especially in drug resistance, we consider one of the first, classical mathematical models of density-dependent population dynamics, Verhulst's logistic growth model \parencite{Verhulst:1838}, which describes the dynamics of a homogeneous population in terms of its net growth rate:
\begin{align}
    \dfrac{d\phi}{dt} = r\Big(1 - \dfrac{\phi}{K} \Big)\phi = r\phi - \dfrac{r}{K}\phi^2. \label{eq:logistic-deterministic}
\end{align}
In Equation \eqref{eq:logistic-deterministic}, $\phi$ denotes population size, $r$ denotes intrinsic \textit{per capita} net growth rate, and $K$ denotes carrying capacity.
The density dependence term $\dfrac{r}{K}\phi^2$ describes the direct or indirect interactions between individuals in the population. The minus ($-$) sign indicates the interactions have a negative \emph{net} effect on the population--in particular, reducing the population size. 
We may consider this negative \emph{net} effect as the difference between a positive effect and a negative effect by introducing a parameter $c \geq 0$:
\begin{align}
    \dfrac{d\phi}{dt} = r\phi - \dfrac{r}{K}\phi^2 &= r\phi + \underbrace{c\dfrac{r}{K}\phi^2}_{\text{cooperation}} - \underbrace{(1+c)\dfrac{r}{K}\phi^2}_{\text{competition}}.
\end{align}
In the context of ecology, we may interpret the term $c\dfrac{r}{K}\phi^2$ as cooperation, the term $(1+c)\dfrac{r}{K}\phi^2$ as competition, and the parameter $c$ as a measure of cooperation. 
In this paper, we consider only competition (i.e. $c=0$). For future work on the cases where $c > 0$, please Section \ref{sect: discussion}. Competitive interactions between individuals can hinder the growth of population size through either the birth process, death process, or some combination of the two. 
However, the formulation in Equation \eqref{eq:logistic-deterministic} leaves the underlying nature of the density dependence unclear.
Density dependence can be manifest in the birth process, death process, or some combination of the two processes.
To disambiguate birth-related vs.~death-related mechanisms, we rewrite the density dependence term with the parameter $\gamma$ as follows:
\begin{align}
    \dfrac{r}{K}\phi^2 = \gamma \dfrac{r}{K}\phi^2 + (1-\gamma)\dfrac{r}{K}\phi^2, \quad 0 \leq \gamma \leq 1.
\end{align}
We interpret the term $\gamma \dfrac{r}{K}\phi^2$ as the reduction in the population's growth rate due to competition-regulated mechanisms affecting the birth process, and $(1-\gamma)\dfrac{r}{K}\phi^2$ as the population's competition-regulated mechanisms affecting the death process.
\\\\
For completion, we also disentangle the intrinsic net growth rate $r\phi$ into birth and death as follows:
\begin{align}
r\phi  = b_0\phi - (b_0 - r)\phi = b_0\phi - d_0\phi, \quad \text{ with } b_0 \geq r > 0 \text{ and } d_0 := b_0 - r \ge 0,
\end{align}
and interpret $b_0$ as the population's intrinsic (low-density) \textit{per capita} birth rate and $d_0$ as the population's intrinsic (low-density) \textit{per capita}  death rate. 
Hence, we parameterize Equation (\ref{eq:logistic-deterministic}) with $\gamma$, $b_0$, and $d_0$ as follows:
\begin{align}
    \dfrac{d\phi}{dt} &= \underbrace{\Bigg(b_0 -  \gamma\dfrac{r}{K}\phi\Bigg)\phi}_{\text{birth}} -  \underbrace{\Bigg(d_0 +  (1-\gamma)\dfrac{r}{K}\phi\Bigg)\phi}_{\text{death}}, \quad 0 \leq \gamma \leq 1, \quad r = b_0 - d_0.
    \label{eq:logistic-deterministic-parameterized}
\end{align}
For fixed $K$, $b_0$, and $d_0$ (or $r$), while different values of $\gamma$ in Equation (\ref{eq:logistic-deterministic-parameterized}) result in equations algebraically equivalent to Equation (\ref{eq:logistic-deterministic}), they describe different density-dependent biological processes.
For example, in ecology, one distinguishes exploitative competition, where limited resources hinder the growth of the populations, from interference competition, where individuals fight against one another \parencite{Jesen:1987}.
The former is manifest in density-dependent birth rates, while the latter leads to density-dependent death rates.
The term $\gamma\dfrac{r}{K}\phi^2$ in Equation (\ref{eq:logistic-deterministic-parameterized}) can be interpreted as the case where individuals have to compete for resources and experience reduced birth due to unfavorable living conditions.
In contrast, the term $(1-\gamma)\dfrac{r}{K}\phi^2$ in Equation (\ref{eq:logistic-deterministic-parameterized}) can be interpreted as the case where interactions between individuals lead to increased death.
Nevertheless, both cases may result in the same net growth rates. 
This example motivates us to ask the following question:
\\\\
\textit{[Q]: In the context of density-dependent population dynamics, how much of a population's change in net growth is through mechanisms affecting birth and how much is through mechanisms affecting death?}\label{main-question}
\\\\
The significance of the answer to this question can also be seen in other contexts.
The Allee effect \parencite{allee1932studies} of density-dependent dynamics (a positive correlation between population density and \textit{per capita} net growth rate)
provides another example. 
Although the Allee effect is typically modeled with cubic growth \parencite{Kanarek:2010} instead of logistic growth, answering question [Q] would contribute to understanding the mechanisms that give rise to the effect. 
Increasing \textit{per capita} net growth rates with increased population density could result from 
increased cooperation or mating among individuals (increased birth rates) or from a reduction in fighting due to  habitat amelioration (decreased death rates) \parencite{Drake:2011}. 
This distinction is important because populations that experience the Allee effect can become extinct if the population sizes fall below the Allee threshold \parencite{Strang:2019}.
Extinction problems are of interest because, for example, we hope to eventually eradicate tumors and harmful bacteria within individual hosts.
Clinically, bactericidal drugs such as penicillin promote cell death, while bacteriostatic drugs such as chloramphenicol, clindamycin, and linezolid inhibit cell division \parencite{Pankey:2004}. \parencite{Lobritz:2015} shows that bactericidal and bacteriostatic drugs affect cellular metabolism differently, and the bacterial metabolic state in turn influences  drug efficacy. 
Identifying ``-cidal'' versus ``-  static'' drugs may help contribute to developing more efficacious drug treatments.
From an evolutionary perspective, \parencite{Frenoy:2018} shows that assuming a zero death rate leads to overestimating bacterial mutation rates under stress, which in turn can lead to incorrect conclusions about the evolution of bacteria under drug treatments. The authors point out that it is important to separately identify birth and death rates.
In another context of evolutionary dynamics, one may compute probability of extinction/escape and mean first-passage time to extinction/escape for cell populations under certain drug treatments such as \parencite{iwasa2003evolutionary, Komarova:2006, foo2010evolution}. 
If we consider evolution as a birth-death process as in \parencite{doebeli2017towards}, computing the probability and mean first-passage time involves separate birth and death rates \parencite{bailey1991elements,norris1998markov,gardiner2009stochastic}, and cell populations with the same net growth rates--but different birth and death rates--can have different extinction/escape probabilities and mean first-passage times.
In fact, \parencite{doebeli2017towards} points out that defining ``fitness'' as net growth rate (difference between birth rate and death rate) loses evolutionary information; instead, we should use separate birth and death rates to measure ``survival of the fittest.'' 
Therefore, the significance of disambiguating birth and death rates underlying a given net growth rate is clear across multiple biological contexts on different scales.
\\\\
In this paper, we aim to answer question [Q] by extracting birth and death rates from  observations of density-dependent population dynamics.
One type of population dynamics information that we can easily observe is population size. 
However, deterministic dynamical models of populations of size $\hat{\phi}$ do not allow us to disentangle birth rate $b_{\hat{\phi}}$ and death rate $d_{\hat{\phi}}$ from net growth rate $(b_{\hat{\phi}}-d_{\hat{\phi}})$, as the transformations $b_{\hat{\phi}} \rightarrow b_{\hat{\phi}}+a_{\hat{\phi}}$ and $d_{\hat{\phi}} \rightarrow d_{\hat{\phi}}+a_{\hat{\phi}}$ leave $(b_{\hat{\phi}} - d_{\hat{\phi}})$ unchanged. 
At a fundamental level, population growth is driven by the birth/division\footnote{Although cells do not give birth to offspring in the biological sense, for the rest of the manuscript, we refer to cell division as birth to be consistent with the birth-death process model we use.} and death of individual cells. 
At this level, cell birth and death are discrete rather than continuous processes, and may involve stochastic elements such as molecular fluctuations in the reactions within individual cells \parencite{Lei:2015}.
Therefore, although the tractability of deterministic population equations has made them attractive as a framework for modeling the growth of pathogenic populations and their responses to therapeutic agents \parencite{Yoon:2018,yoon2021theoretical,Scarborough:2021}, a stochastic modeling framework is more appropriate for the research question we consider. 
Specifically, we consider a birth-death process describing a homogeneous cell population.
We describe density dependence with logistic growth because it is one of the simplest form of density-dependent dynamics and still captures some realistic cell population dynamics such as the dynamics of cancer cells \parencite{gerlee2013model}. Therefore, we will consider a logistic birth-death process model in this paper.
\\\\
The remainder of this manuscript is structured as follows. 
In Section \ref{sect:model}, we describe the mathematical model. 
Then, we describe our direct estimation method in Section \ref{sect:direct-estimation}, where we also validate our method and analyze estimation errors. 
Next, in Section \ref{sect:applications}, we use our direct estimation method to answer question [Q] with a focus on disambiguating autoregulation, drug efficacy, and drug resistance mechanisms. In Section \ref{sect:likelihood}, we present the likelihood-based inference approach that deals with small sample sizes.
Finally, in Section \ref{sect: discussion} we compare our approach to related existing methods \parencite{Crawford:2014,Liu:2018,Ferlic:2019}, and discuss future directions.
\section{Mathematical Model} \label{sect:model}
\begin{figure}[H]
  \centering
  \includegraphics[scale=0.85]{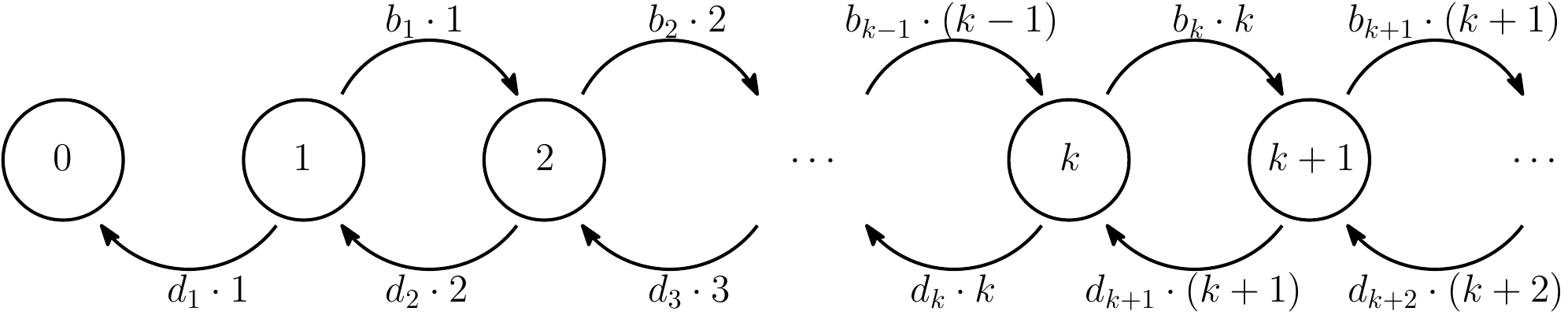}
  \caption{\textbf{Schematic representation of our birth-death process model.} The \textit{per capita} birth rate $b_k$, and \textit{per capita} death rate $d_k$ depend on cell number $N = k$ with $k = 0, 1, \ldots$. State $N = k$ transitions to state $N = k+1$ at rate $b_k\cdot k$ and transitions to state $N = k-1$ at rate $d_k\cdot k$. At state $N = 0$, the system cannot transition to state $N = 1$, because there is no individual to give birth.}.
  \label{fig:birth-death-process}
\end{figure}

We consider systems of homogeneous cells described by a birth-death process, that is, a discrete-state continuous-time Markov chain tracking the number of individual cells $N(t)$ in the system over time $t$, with state transitions comprising either ``birth'' ($N\to N+1$) or ``death'' ($N\to N-1$), as shown in Figure \ref{fig:birth-death-process}.
In linear birth-death processes, \textit{per capita} birth and death rates are constants that do not depend on $N$. 
In contrast, here we consider birth-death processes whose \textit{per capita} birth and death rates depend on $N$, in order to incorporate density-dependent population dynamics. 
Specifically, motivated by Equation \eqref{eq:logistic-deterministic-parameterized}, we define the \textit{per capita} birth rate $b_N$ and death rate $d_N$ in our model as follows:
\begin{align}
    b_{N} &= \max\left\{b_0 - \gamma \dfrac{r}{K}N,0\right\}\label{eq:birth-rate-definition},\\
    d_{N} &= d_0 + (1-\gamma)\dfrac{r}{K}N, \label{eq:death rate-definition}
\end{align}
where $b_0 > 0$ and $d_0 \geq 0$ are intrinsic (low-density) \textit{per capita} birth and death rates respectively, $r = b_0 - d_0 \geq 0$ is the intrinsic (low-density) \textit{per capita} net growth rate, $K > 0$ is the population's carrying capacity, and $\gamma \in [0,1]$ determines the extent to which the nonlinear or density-dependent dynamics arises from the \textit{per capita} birth versus death rates. When $\gamma = 0$, the birth process is density-independent; all density dependence lies in the death process.
Conversely, when $\gamma = 1$, the density-dependent dynamics is fully contained in the birth process.
When $0 < \gamma < 1$, the density-dependent dynamics is split between birth and death. We use the $\max$ function in Definition (\ref{eq:birth-rate-definition}) to ensure $b_N$ is nonnegative. The total birth and death rates of the population are $b_NN$ and $d_NN$. 
\noindent
For a single-species birth-death process of this form, with $d_1>0$ ($d_1$ is the death rate when $N = 1$) and no immigration, it is well known that the unique stationary probability distribution gives $N(t)\to0$ as $t\to\infty$ with probability one \parencite{Allen:2010}.  
Rather than concern ourselves with the long-term behavior, here we are interested in answering question [Q] by estimating $b_N$ and $d_N$.  
Therefore we will focus on the analysis of transient population behavior rather than long-time, asymptotic behavior.

\section{Direct Estimation of Birth and Death Rates}\label{sect:direct-estimation}
In this section, we describe our method of estimating the birth and death rates of cell populations that follow the logistic birth-death process model described in Section \ref{sect:model}. We would like to disambiguate different pairs of birth and death rates for the same observed mean change in population size. 

\subsection{Mathematical Derivation}
\label{ssec:math}
Let $N(t)\ge 0$ be an integer-valued random variable representing the number of cells at time $t$. 
We consider a small time increment  $\Delta t$, within which each cell can either divide (i.e.~one cell is replaced by two cells), die (i.e.~one cell disappears and is not replaced), or stay the same (i.e.~there is still one cell). 
Focusing on a single timestep, let $(\Delta N_+|N,\Delta t)$ and $(\Delta N_-|N,\Delta t)$ be two random variables representing the numbers of cells gained and lost, respectively, from an initial population of $N$ cells, after a period of time $\Delta t$. 
The number of cells that neither die nor divide is thus equal to $N-\Delta N_+-\Delta N_-$.
Although the two random variables $(\Delta N_+|N,\Delta t)$ and $(\Delta N_-|N,\Delta t)$ are not strictly independent (as one cell cannot both die and reproduce at the same time), we work in a regime in which the correlation between them is small enough to be neglected.
Among $N$ cells, $\Delta N_+$ cells are ``chosen'' to divide and $\Delta N_-$ cells are ``chosen'' to die. 
On a time interval of length $\Delta t$, the probabilities that  a cell divides and dies are $b_N\Delta t + o(\Delta t)$ and $d_N\Delta t + o(\Delta t)$ respectively.\footnote{We adopt the standard convention $\dfrac{o(\Delta t)}{\Delta t}\to 0$ as $\Delta t\to 0$.}
For convenience, we will omit the $o(\Delta t)$ correction where possible without introducing inaccuracies.
The random variables $(\Delta N_+|N, \Delta t)$ and  $(\Delta N_-|N, \Delta t)$ are binomially distributed. In particular, 
\begin{align}
(\Delta N_+|N,\Delta t) &\sim \text{ Binomial}(N,b_N\Delta t) \text{ with mean } Nb_N\Delta t \text{ and variance } Nb_N\Delta t(1 - b_N\Delta t),\\
(\Delta N_-|N,\Delta t)&\sim \text{ Binomial}(N,d_N\Delta t) \text{ with mean } Nd_N\Delta t \text{ and variance } Nd_N\Delta t(1 - d_N\Delta t).
\end{align}
Define a random variable $(\Delta N|N,\Delta t)$ to be the net change in population size from $N$ cells after a period of time $\Delta t$, i.e. $(\Delta N|N,\Delta t) = (\Delta N_+|N,\Delta t) - (\Delta N_-|N,\Delta t)$.
Typically, experimental or clinical measurements reflect only the net change $(\Delta N|N,\Delta t)$ rather than the increase $(\Delta N_+|N,\Delta t)$ or decrease $(\Delta N_-|N,\Delta t)$ separately.
Because $(\Delta N_+|N,\Delta t)$ and $(\Delta N_-|N,\Delta t)$ are approximately independent, for sufficiently small $\Delta t$, we have
\begin{align}
    \E[\Delta N|N,\Delta t] &= \E[\Delta N_+|N,\Delta t]-\E[\Delta N_-|N,\Delta t] = Nb_N\Delta t - Nd_N\Delta t = (b_N-d_N)N\Delta t,\\
    \V[\Delta N|N,\Delta t] &\approx \V[\Delta N_+|N,\Delta t]+\V[\Delta N_-|N,\Delta t] = Nb_N\Delta t(1 - b_N\Delta t) + Nd_N\Delta t(1 - d_N\Delta t)\\
    &= Nb_N\Delta t + Nd_N\Delta t + O(\Delta t^2)\\
    &\approx (b_N+d_N)N\Delta t.
\end{align}
Therefore, to estimate birth and death rates $b_NN$ and $d_NN$, we solve the linear system:
\begin{align}
    \label{eq:birth-death-linear-system-analytic}
    (b_N-d_N)N = \dfrac{\E[\Delta N|N,\Delta t]}{\Delta t} \text{  and  } (b_N+d_N)N = \dfrac{ \V[\Delta N|N,\Delta t]}{\Delta t}.
\end{align}
In Section \ref{sect: direct-method}, we discuss how we obtain approximations to $\E[\Delta N|N,\Delta t]$ and $\V[\Delta N|N,\Delta t]$ from discretely sampled finite time series.

\subsection{Data Simulation} \label{sect:data-simulation}
To validate our method, we use simulated \textit{in silico} data. 
While our underlying model is time-continuous, in experimental and clinical settings, one can only observe cell numbers at discrete time points. 
In order to efficiently generate an ensemble of trajectories of the birth-death process, we construct a $\tau$-leaping approximation \parencite{gillespie2001approximate} as follows.
\\\\
Given $N(t)$ individuals at time $t$, we approximate the number of individuals after a short time interval $\Delta t$ as
\begin{align}
    N(t+\Delta t) \approx N(t) + \Delta N_+(t) - \Delta N_-(t),
\end{align}
where $\Delta N_+\sim\text{Binomial}\Big(N(t),b_{N(t)}\Delta t\Big)$ and $\Delta N_-\sim\text{Binomial}\Big(N(t),d_{N(t)}\Delta t\Big)$ representing the number of cells added to and lost from the system after a period of time $\Delta t$.
We approximate $\Delta N_+$ and $\Delta N_-$ as if they were independent random variables; see discussion in Section \ref{ssec:math}.
When $N(t)$ is sufficiently large, we approximate the binomial distributions with Gaussian distributions that have the same means and variances as the binomial distributions. Because our discrete-state process in Section \ref{sect:model} is now approximated with a continuous-state process, we replace $N(t)$ with a different notation, $X(t)$, to make this approximation clear. We have
\begin{align}
    \Delta X(t)_+ &\sim \text{Normal}\Bigg(X(t)b_{X(t)}\Delta t, X(t)b_{X(t)} \Delta t\Big(1-b_{X(t)}\Delta t\Big)   \Bigg),\\
    \Delta X(t)_- &\sim \text{Normal}\Bigg(X(t)d_{X(t)}\Delta t, X(t)d_{X(t)} \Delta t\Big(1-d_{X(t)}\Delta t\Big) \Bigg).
\end{align}
Thus, the net change in number of cell after a timestep $\Delta t$ is
\begin{align}
    X(t+\Delta t) - X(t) \approx& X(t) b_{X(t)} \Delta t + \Delta W_+(t)\sqrt{X(t)b_{X(t)} \Delta t\Big(1-b_{X(t)} \Delta t\Big)}\\
    &- X(t) d_{X(t)} \Delta t - \Delta W_-(t)\sqrt{X(t)d_{X(t)} \Delta t\Big(1-d_{X(t)}\Delta t\Big)} \nonumber\\
    =& \Big(b_{X(t)} - d_{X(t)}\Big)X(t)\Delta t + \sqrt{\Big(b_{X(t)} + d_{X(t)}\Big)X(t)}\Delta W(t), \label{eq:tau-leaping-step}
\end{align}
where $\Delta W_\pm$ are independent Wiener process increments, and $\Delta W$ is a Wiener process increment derived from a linear combination of the $\Delta W_\pm$.
Equation \eqref{eq:tau-leaping-step} is the $\tau$-leaping approximation used in our data simulation, which is analogous to the forward Euler algorithm in the deterministic setting. 
Taking the limit $\Delta t \rightarrow dt$, we obtain a version of our population model as a continuous-time Langevin stochastic differential equation
\begin{align}
    dX(t) &=  \Big(b_{X(t)} - d_{X(t)} \Big)X(t)dt + \sqrt{\Big(b_{X(t)} + d_{X(t)} \Big)X(t)}\, dW(t). \label{eq:stochastic-differential-equation}
\end{align}
where $dW(t)$ is delta-correlated white noise satisfying $\langle dW(t)dW(t')\rangle=\delta(t-t').$
We use Equation \eqref{eq:stochastic-differential-equation} under the Ito interpretation.  

\subsection{Direct Estimation} \label{sect: direct-method}
We conduct $S$ experiments to collect an ensemble of $S$ cell number time series and obtain the following dataset
\begin{align}
\mathcal{D} &= \Big\{\underbrace{[X(t^1_0),\ldots,X(t^1_{T_1})]^T}_{\text{time series 1}}, \ldots, \underbrace{[X(t^s_0),\ldots,X(t^s_{T_1})]^T}_{\text{time series s}}, \ldots, \underbrace{[X(t^S_0),\ldots,X(t^S_{T_S})]^T}_{\text{time series $S$}}\Big\},
\end{align}
each of which has $T_s+1$ data points, $s = 1, \ldots, S$. 
Note that we use the notation $X$ to represent data for the continuous random cell number under a Gaussian approximation, as discussed in Section \ref{sect:data-simulation}.
We use $\tau$-leaping simulation so that for all the time series indices $s \in \{1, \ldots, S\}$ and all the time point indices $j \in \{0,
\ldots,T_s-1\}$, the difference $t^s_{j+1}-t^s_{j}$ is equal to $\Delta t$, which is independent of $s$ and $j$, which is consistent with the format of the dataset produced from the EVolutionary biorEactor (EVE) experiments in our laboratory \parencite{gopalakrishnan2020low}.
In our simulation, for all time series $s = 1, \ldots, S$, we choose $t^s_0$ to be equal to $t_0$ and $T_s$ to be equal to $T$ so that each time series has the same number of data points as the others.
\\\\
In order to obtain the statistics of the cell number increments, conditioned on the population size, we consider the truncated dataset
\begin{align}
\mathcal{D}_{-1} &= \Big\{\underbrace{[X(t^1_0),\ldots,X(t^1_{T_1-1})]^T}_{\text{time series 1}}, \ldots, \underbrace{[X(t^s_0),\ldots,X(t^s_{T_s-1})]^T}_{\text{time series s}}, \ldots, \underbrace{[X(t^S_0),\ldots,X(t^S_{T_S-1})]^T}_{\text{time series $S$}}\Big\},
\end{align}
in which we omit the last element of each of the $S$ time series in $\mathcal{D}$.
We put all the data points in $\mathcal{D}_{-1}$ across the whole ensemble of trajectories into bins along the population axis. 
Denote the bin size as $\eta$.
The left end point $X_k$ of the $k$th bin $[X_k, X_k + \eta)$, $k = 1, 2, \ldots, k_{\max}$, is equal to $X_k := X_{\min} + (k-1)\eta$, where $X_{\min}$ is the smallest value of cell number across the whole dataset $\mathcal{D}_{-1}$.
The total number of bins $k_{\max} \in \mathbb{Z}^+$ is equal to $\Big\lceil\dfrac{X_{\max} - X_{\min}}{\eta} \Big\rceil$, where $X_{\max}$ is the largest value of cell number across the whole dataset $\mathcal{D}_{-1}$.
\\\\
Denote $\hat{S}_k$ as the number of elements in the $k$th bin $[X_k,X_k + \eta)$.
Our method requires a sufficiently large bin size so that the bins have at least two entries in $\mathcal{D}_{-1}$ in order to compute the variances of the cell number increments.
For each point $X=X_k+\eta_i$ in the $k$th bin, $0 \leq \eta_i < \eta$, $i=1,2,\ldots,\hat{S}_k$, let $\Delta X_{ki}$ be the subsequent increment in $X$, i.e.~$\Delta X_{ki}=X(t_*+\Delta t) - X(t_*)$, where $t_*$ is the time corresponding to $X=X_k+\eta_i$.  
For each $k$th bin, $k = 1, 2, \ldots, k_{\max}$, we compute the empirical mean and variance of the cell number increments $\{\Delta X_{ki}\}_{i=1}^{\hat{S}_k}$ , 
and use these statistics (e.g.~mean and variance) 
to estimate the birth and death rates corresponding to the population size $X = X_k + \dfrac{\eta}{2}$.

\subsection{Validation and Error Analysis}\label{sect:valid-error-analysis}
\begin{figure}[H]
  \centering
  \includegraphics[scale=0.85]{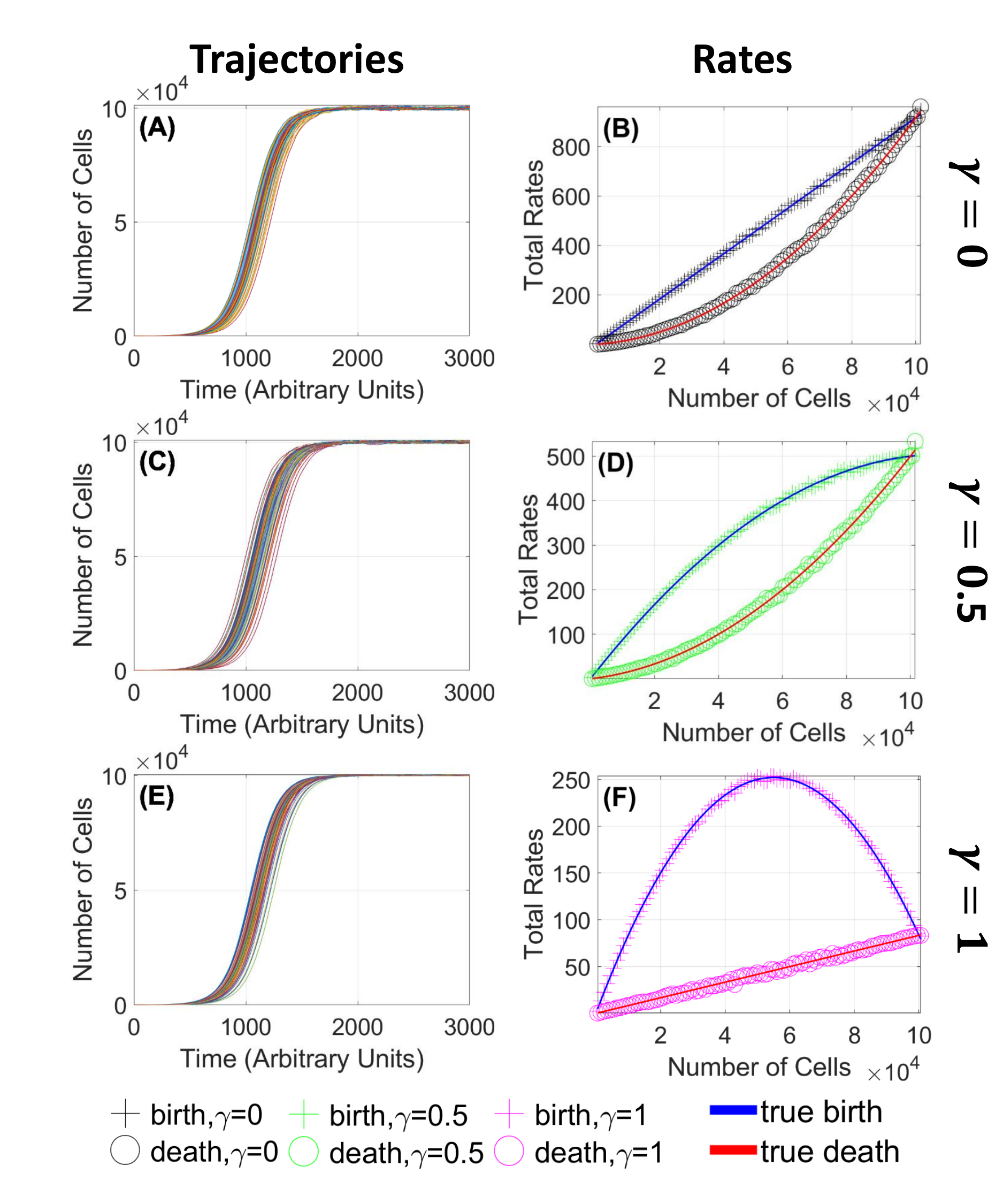}
  \caption{\textbf{Agreement of estimated and true birth and death rates validates the direct estimation method.}
  \textbf{(A, C, E)}: Time series ensembles simulated using the $\tau$-leaping approximation for the cases $\gamma = 0$ \textbf{(A)}, $\gamma = 0.5$ \textbf{(C)}, and $\gamma = 1$ \textbf{(E)} respectively. Each figure shows $S=100$ trials. The estimated rates are computed using a bin size of $\eta=10^3$.
  Carrying capacity $K=10^5$ cells; low-density growth rate $r=1/120$ (arbitrary time units); for other parameters see Table \ref{table:simulation-parameters}.
  \textbf{(B, D, F)}: Estimated and true birth and death rates, as functions of population size. 
  Blue line: true birth rate. 
  Red line: true death rate. 
  Plus signs $(+)$ denote estimated birth rates; circles $(\circ)$ denote estimated death rates.
  Throughout the paper we will use distinct colors to denote values of $\gamma$. \textbf{(B)} Black: $\gamma=0$;  \textbf{(D)} Green: $\gamma=0.5$; \textbf{(E)} Magenta: $\gamma=1.0$. We observe that the estimated birth and death rates are well-aligned with the true birth and death rates used to simulate the trajectories in \textbf{(A)}, \textbf{(C)}, and \textbf{(E)}.} 
  \label{fig:compare-estimated-true-rates}
\end{figure}

\begin{figure}[H]
  \centering
  \includegraphics[scale=0.85]{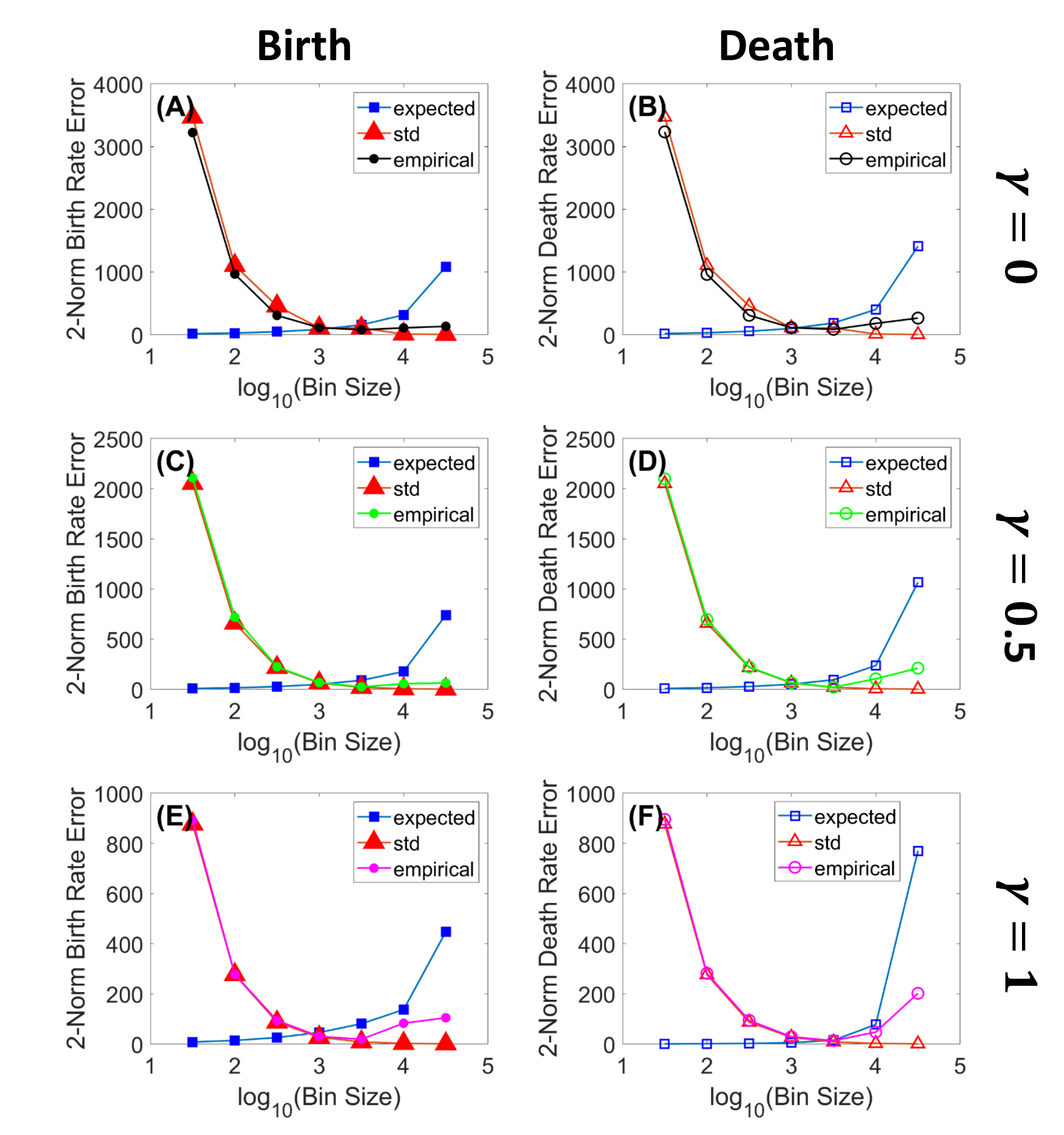}
  \caption{\textbf{Intermediate bin sizes give optimal estimation performance.} We plot the $l_2$-norm (over all bins) errors in estimating birth rate (left column) and death rate (right column) as functions of bin size $\eta$ for carrying capacity $K = 10^5$. Squares ($\square$)  denote expected values of errors; triangles ($\bigtriangleup$) denote standard deviations of errors; circles (◦) denote empirical errors using data from a simulation of $S=100$ cell number trajectories. \textbf{(A, C, E)}: errors in estimating birth rates. \textbf{(B, E, F)}: errors in estimating death rates. \textbf{(A, B)}: $\gamma = 0$ (black color); \textbf{(C, D)}: $\gamma = 0.5$ (green color); \textbf{(E, F)}: $\gamma = 0.5$ (magenta color). We observe that as the bin size $\eta$ increases, the expected errors increase, the theoretical variances/standard deviations of the errors decreases, and the sample errors balance between the expected values and variances and have convex quadratic shapes.} 
  \label{fig:erroranalysis-errorbirthdeath}
\end{figure}

We validate our method by comparing estimated rates with ``true'' rates that are used to generate the simulated data.
Specifically, we simulate $S=100$ cell number trajectories, using a numerically efficient $\tau$-leaping approximation described in Section \ref{sect:data-simulation},  
and estimate birth and death rates using Equations \eqref{eq:birth-death-linear-system-analytic} and the method described in Section \ref{sect: direct-method}. 
Figure \ref{fig:compare-estimated-true-rates} shows that the estimated and true rates are well-aligned. 
Figure \ref{fig:compare-estimated-true-rates} \textbf{(A, C, E)} shows an ensemble of $S=100$ independent realizations of the logistic birth-death process formulated in Section \ref{sect:model} for three scenarios: $\gamma = 0$ (black), $\gamma = 0.5$ (green), and $\gamma = 1$ (magenta), respectively, simulated using the $\tau$-leaping method with the initial condition $N(t_0) = 10$ and the model parameter values in Table \ref{table:simulation-parameters}, over a time period of length 3000 (arbitrary units) and timestep $\Delta t = 1/30$. 
\noindent
Figure \ref{fig:compare-estimated-true-rates} \textbf{(B, D, F)} shows the corresponding true and estimated birth and death rates, using a bin size of $\eta=10^3$.
The true birth and death rates are solid blue and red lines respectively. 
Plus signs $(+)$ denote estimated birth rates, and circles $(\circ)$ denote estimated death rates. We observe that the true and estimated rates are well-aligned.
\noindent
Using the discretization described in Section \ref{sect: direct-method}, we estimate birth and death rates via the empirical mean $\Big\langle\Delta N \Big| N = N_k + \eta_i, 0 \leq \eta_i < \eta, \hat{S}_k\Big\rangle$ and empirical variance $\sigma^2\Big[\Delta N \Big| N = N_k + \eta_i, 0 \leq \eta_i < \eta,\hat{S}_k\Big]$ obtained from an ensemble of $S=100$ simulated trajectories.
To quantify the accuracy of our method, we define the error $\mathcal{E}_{k\text{birth}}$ in estimating the birth rate corresponding to population size 
$N = N_k + \dfrac{\eta}{2}$, and the error $\mathcal{E}_{k\text{death}}$ in estimating the death rate corresponding to $N = N_k + \dfrac{\eta}{2}$ as follows:
\begin{align}
    \mathcal{E}_{k\text{birth}} :=& \dfrac{\E\Big[\Delta N \Big| N = N_k + \dfrac{\eta}{2}\Big] + \V\Big[\Delta N \Big| N = N_k + \dfrac{\eta}{2}\Big]}{2\Delta t}\\
    &- \dfrac{\Big\langle\Delta N \Big| N = N_k + \eta_i, 0 \leq \eta_i < \eta, \hat{S}_k\Big\rangle + \sigma^2\Big[\Delta N \Big| N = N_k + \eta_i, 0 \leq \eta_i < \eta,\hat{S}_k\Big]}{2\Delta t},\\
    \mathcal{E}_{k\text{death}} :=& \dfrac{\V\Big[\Delta N \Big| N = N_k + \dfrac{\eta}{2}\Big]-\E\Big[\Delta N \Big| N = N_k + \dfrac{\eta}{2}\Big]}{2\Delta t}\\
    &-  \dfrac{\sigma^2\Big[\Delta N \Big| N = N_k + \eta_i, 0 \leq \eta_i < \eta,\hat{S}_k\Big]-\Big\langle\Delta N \Big| N = N_k + \eta_i, 0 \leq \eta_i < \eta, \hat{S}_k\Big\rangle}{2\Delta t}.\\
\end{align}
\noindent
Under the assumption that the samples $\eta_i$ are iid uniformly distributed on $[0,\eta)$, the theoretical means and variances of the errors $\mathcal{E}_{k\text{birth}}$ and $\mathcal{E}_{k\text{death}}$ are equal to
\begin{align}
    \E\Big[\mathcal{E}_{k\text{birth}}\Big] =& \dfrac{\E\Bigg[\E\Big[\Delta N \Big| N = N_k + \dfrac{\eta}{2}\Big]\Bigg] + \E\Bigg[\V\Big[\Delta N \Big| N = N_k + \dfrac{\eta}{2}\Big]\Bigg]}{2\Delta t}\\
    &- \dfrac{\E\Bigg[\Big\langle\Delta N \Big| N = N_k + \eta_i, 0 \leq \eta_i < \eta, \hat{S}_k\Big\rangle\Bigg] + \E\Bigg[\sigma^2\Big[\Delta N \Big| N = N_k + \eta_i, 0 \leq \eta_i < \eta,\hat{S}_k\Big]\Bigg]}{2\Delta t}\\
    =& \dfrac{\E\Big[\Delta N \Big| N = N_k + \dfrac{\eta}{2}\Big] - \E\Big[\Delta N \Big| N = N_k + U, U\sim \text{Unif}[0,\eta)\Big]}{2\Delta t}\\
    &+ \dfrac{\V\Big[\Delta N \Big| N = N_k + \dfrac{\eta}{2}\Big] - \V\Big[\Delta N \Big| N = N_k + U, U\sim \text{Unif}[0,\eta)\Big]}{2\Delta t},
\end{align}
\begin{align}
    \E\Big[\mathcal{E}_{k\text{death}}\Big] =& \dfrac{\E\Bigg[\V\Big[\Delta N \Big| N = N_k + \dfrac{\eta}{2}\Big]\Bigg]-\E\Bigg[\E\Big[\Delta N \Big| N = N_k + \dfrac{\eta}{2}\Big]\Bigg]}{2\Delta t}\\
    &- \dfrac{\E\Bigg[\Big\langle\Delta N \Big| N = N_k + \eta_i, 0 \leq \eta_i < \eta, \hat{S}_k\Big\rangle\Bigg] + \E\Bigg[\sigma^2\Big[\Delta N \Big| N = N_k + \eta_i, 0 \leq \eta_i < \eta,\hat{S}_k\Big]\Bigg]}{2\Delta t}\\
    =& \dfrac{\E\Big[\Delta N \Big| N = N_k + \dfrac{\eta}{2}\Big] - \E\Big[\Delta N \Big| N = N_k + U, U\sim \text{Unif}[0,\eta)\Big]}{2\Delta t}\\
    &+ \dfrac{\V\Big[\Delta N \Big| N = N_k + \dfrac{\eta}{2}\Big] - \V\Big[\Delta N \Big| N = N_k + U, U\sim \text{Unif}[0,\eta)\Big]}{2\Delta t}.
\end{align}
Similarly,
\begin{align}
    &\V\Big[\mathcal{E}_{k\text{birth}}\Big] = \V\Big[\mathcal{E}_{k\text{death}}\Big]\\
    &= \dfrac{\V\Bigg[\Big\langle\Delta N \Big| N = N_k + \eta_i, 0 \leq \eta_i < \eta, \hat{S}_k\Big\rangle\Bigg] + \V\Bigg[\sigma^2\Big[\Delta N \Big| N = N_k + \eta_i, 0 \leq \eta_i < \eta,\hat{S}_k\Big]\Bigg]}{4\Delta t^2}\\
    &= \dfrac{\dfrac{\V\Big[\Delta N \Big| N = N_k + \dfrac{\eta}{2}\Big]}{\hat{S}_k} + \dfrac{2\Big(\V\Big[\Delta N \Big| N = N_k + U, U\sim \text{Unif}[0,\eta)\Big]\Big)^2}{\hat{S}_k-1}}{4\Delta t^2}.
\end{align}
\noindent
We compute $\E\Big[\mathcal{E}_{k\text{birth}}\Big]$, $\E\Big[\mathcal{E}_{k\text{death}}\Big]$, $\V\Big[\mathcal{E}_{k\text{birth}}\Big]$, and $\E\Big[\mathcal{E}_{k\text{death}}\Big]$ as functions of bin size in Appendix \ref{appendix:appx-error-analysis-direct-method}. We then compute the 2-norm of the theoretical means and standard deviations (i.e. square roots of the variances) over all $k$ to obtain the plots in Figure \ref{fig:erroranalysis-errorbirthdeath}. 
\noindent
We observe that as the bin size $\eta$ increases, the expected errors increase, the theoretical variances (or standard deviations) of the errors decreases, and the empirical errors (computed using data from a simulation of $S = 100$ cell number trajectories) balance between the expected values and variances (or standard deviations), as shown in Figure~\ref{fig:erroranalysis-errorbirthdeath}.
The expected values of errors reflect the differences between $\Delta N$ at the midpoint $\Big(N = N_k + \dfrac{\eta}{2}\Big)$ and $\Delta N$ at multiple points $\Big(N = N_k + \eta_i, 0 \leq \eta_i < \eta\Big)$.
The smaller the bin size, the closer multiple points are to the midpoint, so the error is smaller. 
However, if the bin is too small, then there are 
too few samples to accurately estimate theoretical statistics with empirical statistics.
The theoretical variances of errors involves sample sizes; the bigger the bin size, the more samples we have. 
These two competing effects of bin size result in the empirical errors being intermediate values between the two theoretical statistics (expected values and variances) of the estimation errors. 
This ``Goldilocks principle" is an example of the bias/variance tradeoff common in many estimation problems.

\section{Inferring Underlying Mechanisms of Autoregulation, Drug Efficacy, and Drug Resistance}\label{sect:applications}
In this section, we apply our direct estimation method (Section \ref{sect:direct-estimation}) to shed light on drug resistance mechanisms of pathogenic cell populations (e.g.~malignant tumors or harmful bacteria) by disambiguating whether the mechanisms involve  the birth process, the death process, or both processes.
We consider the scenario where a homogeneous pathogenic cell population grows to its carrying capacity, then is treated with a drug that reduces its carrying capacity, and then overcomes the drug effect to regain  its original carrying capacity. 
Within this scenario, we use ``drug resistance'' to refer to the pathogenic population's recovery of its original carrying capacity. 
(For a discussion of different perspectives on drug resistance, please refer to Section \ref{sect: discussion}). 
We divide our analysis into three stages: (1) auto-regulated growth, (2) drug treatment, and (3) drug resistance. 
The autoregulation stage occurs before the drug treatment stage; during this stage, the cells regulate themselves in such a way that their  growth saturates at a given carrying capacity. 
Such regulation can be due to direct or indirect cell-to-cell interactions, such as exploitation or interference competition.
During the drug treatment stage, the cells are regulated by an applied drug, which reduces the population's carrying capacity.
The reduced carrying capacity may result either by increasing the density-dependent death rate (``-cidal'' effect) or decreasing the density-dependent birth rate (``-static'' effect), or both.
Finally, in the drug resistance stage, after having been treated with either a ``-cidal'' or ``-static'' drug, the cell population fights back and regains to its original carrying capacity by either decreasing its density-dependent death rate or by increasing its density-dependent birth rate. 
In each of these stages,  changes in either birth or death rates could result in the same observed net dynamics. 
It is important to disambiguate the underlying mechanisms, to appropriately design optimal treatments with the goal of eventually eradicating the pathogens (i.e.~reducing their sizes to zero). 
\subsection{Stage 1: Autoregulation}\label{sect:auto-regulation}
\begin{figure}[H]
  \centering
  \includegraphics[scale=0.85]{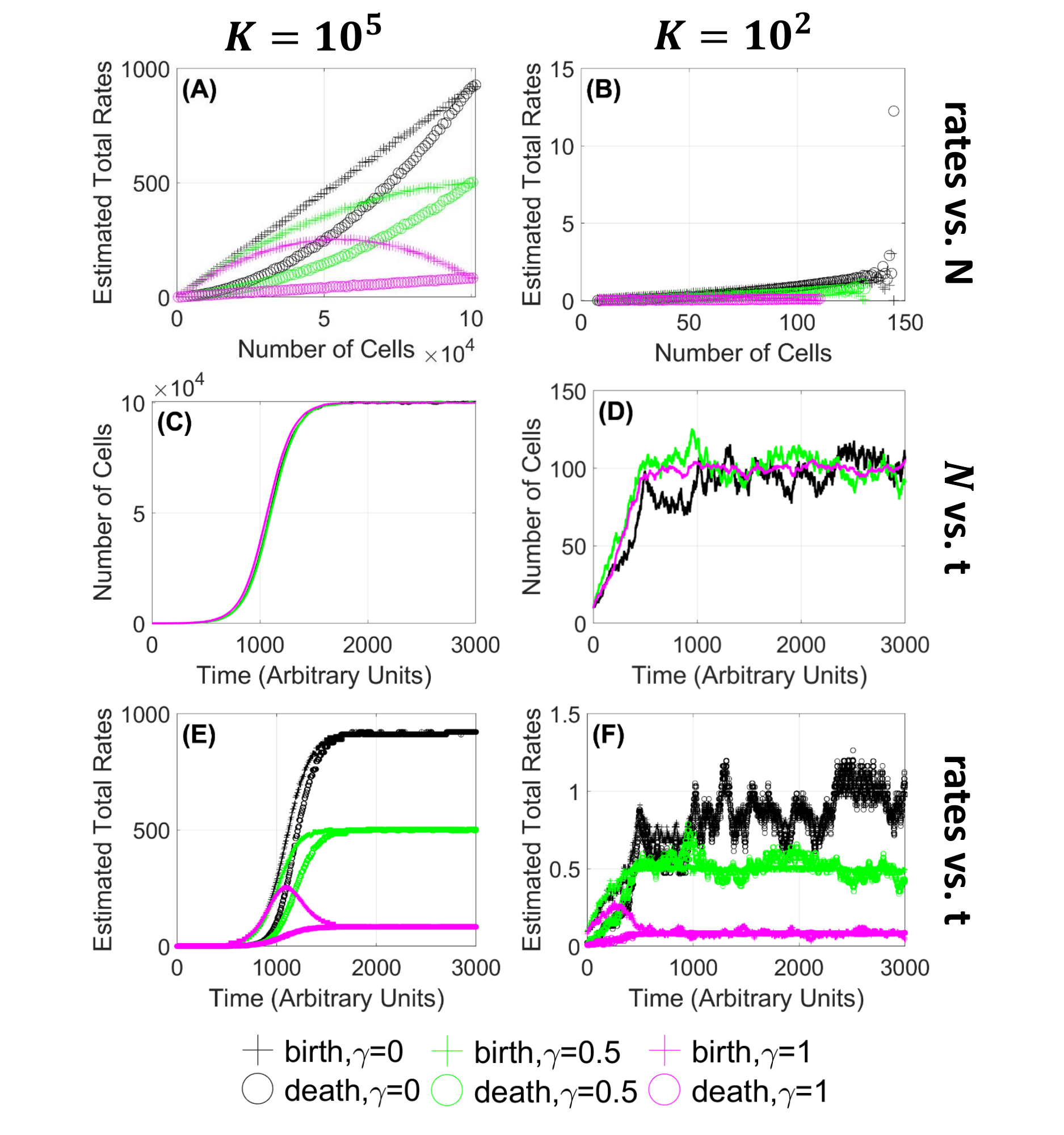}
  \caption{\textbf{Underlying autoregulation mechanisms are distinguished by separately identified birth and death rates, not necessarily by net changes in total population size.}  Plots showing that cell populations with the same net growth rate and carrying capacity grow to the carrying capacity under different density-dependent mechanisms, although the observed dynamics (shown in (C) and (D)) look indistinguishable. Noise levels are more visible for smaller carrying capacities due to smaller scales.
  \textbf{(A, C, E)}: logistic birth-death processes with carrying capacity $K = 10^5$. \textbf{(B, D, F)}: logistic birth-death processes with carrying capacity $K = 10^2$. \textbf{(A-F)}: black curves correspond to the scenario $\gamma = 0$; green curves correspond to the scenario $\gamma = 0.5$; magenta curves correspond to the scenario $\gamma = 1$. \textbf{(A, B)}: estimated birth and death rates for three scenarios using the direct estimation method with an ensemble of 100 trajectories. \textbf{(C, D)}: one selected trajectory for each scenario. \textbf{(E, F)}: estimated birth and death rates throughout time corresponding to the trajectories in \textbf{(C)} and \textbf{(D)}. Plus signs $(+)$ denote estimated birth rates; circles $(\circ)$ denote estimated death rates.}
  \label{fig:selfreg-3cases}
\end{figure}

Cell populations with the same mean net growth rate can grow and reach their carrying capacities through different mechanisms: density-dependent birth dynamics, density-dependent death dynamics, or some combination of the two. 
The differences between theses scenarios are characterized by different values of the density dependence parameter, $\gamma$, in the model described in Section \ref{sect:model}.
We demonstrate this variety with three scenarios:
\begin{itemize}
\item[(I)] Density dependence occurs only in the \textit{per capita} death rate, while the \textit{per capita} birth rate is density independent ($\gamma = 0)$. 
In this case, the \textit{per capita} birth rate is $(b_N)_{\text{original}} = b_0$ and the \textit{per capita} death rate is $(d_N)_{\text{original}} = d_0 + \dfrac{r}{K}N$. 
The plots corresponding to scenario (I) in all the figures in this paper are represented by the color black.
\item[(II)] Density dependence occurs in both the \textit{per capita} birth and death rates ($\gamma = 0.5$). 
In this case, the \textit{per capita} birth rate is $(b_N)_{\text{original}} = \max\left\{b_0 -0.5\dfrac{r}{K}N,0\right\}$ and the \textit{per capita} death rate is $(d_N)_{\text{original}} = d_0 + 0.5\dfrac{r}{K}N$.
The plots corresponding to scenario (II) in all the figures in this paper are represented by the color green.
\item[(III)] Density dependence occurs only in the \textit{per capita} birth rate, while the \textit{per capita} death rate is density-independent ($\gamma = 1$). 
In this case, the \textit{per capita} birth rate is $(b_N)_{\text{original}} = \max\left\{b_0 - \dfrac{r}{K}N,0\right\}$ and the \textit{per capita} death rate is $(d_N)_{\text{original}} = d_0$.
The plots corresponding to scenario (III) in all the figures in this paper are represented by the color magenta.
\end{itemize}
Recall that the random variable $N(t)$ represents the number of cells at time $t$ in the logistic birth-death process described in Section \ref{sect:model}. Similarly, the parameters $b_0$, $d_0$, $r$, $K$, and $\gamma$ are the same as those described in Section \ref{sect:model}.
Scenarios (I-III) have the same net growth rate, $(b_N)_{\text{original}} - (d_N)_{\text{original}}$, but different magnitudes of birth and death rates. 
Scenario (I) represents a situation in which the carrying capacity of the population arises through an increase in the \textit{per capita} death rate with population density. 
Such a scenario could arise, for example, when competition is mediated through cell-to-cell interactions such as predation or other conspecific lethal interactions.
Scenario (III), in contrast, represents a situation in which the \textit{per capita} death rate remains constant with increasing population size, but the \textit{per capita} birth (cell division) rate declines.  
Such a scenario could arise, for example, when competition is mediated by accumulation of waste products or competition for food resources that slow cell division.\footnote{Resources depletion can also increase death rates. In this paper, we neglect such effect.}
Scenario (II), intermediate between (I) and (III), represents a combination of such density-dependent mechanisms.\\\\
Figure \ref{fig:selfreg-3cases} shows how our direct estimation method can disambiguate the three autoregulation scenarios (I), (II), and (III).
We simulate 100 trajectories of the cell population under each scenario with an initial population $N(t_0) = 10$, and the parameter values in Table \ref{table:simulation-parameters}, except the carrying capacity value.
In addition to using $K = 10^5$ for carrying capacity (Figure \ref{fig:selfreg-3cases} \textbf{(A)}, \textbf{(C)}, and \textbf{(E)}), we also simulate the population with a carrying capacity $K = 10^2$ (Figure \ref{fig:selfreg-3cases} \textbf{(B)}, \textbf{(D)}, and \textbf{(F)}).
We demonstrate that when the carrying capacity is small (e.g.~$10^2$), it is easier to see the noise levels than when the carrying capacity is large (e.g.~$10^5$), as seen in Figure \ref{fig:selfreg-3cases} \textbf{(C)} and \textbf{(D)}, because the fluctuations are larger relative to the mean population.
After simulating an ensemble of cell number trajectories, we estimate birth and death rates from that ensemble of trajectories using the method given in section \ref{sect: direct-method}, as shown in Figure \ref{fig:selfreg-3cases} \textbf{(A)} and \textbf{(B)}.
Then, we randomly select one trajectory $N(t)$ from those 100 trajectories, as shown in Figure \ref{fig:selfreg-3cases} \textbf{(C)} and \textbf{(D)}, and plot birth and death rates as functions of time, as shown in Figure \ref{fig:selfreg-3cases} \textbf{(E)} and \textbf{(F)}.
The birth rate $b_{N(t)}$ and death rate $d_{N(t)}$ as functions of time are calculated by treating the rates as composite functions of the cell number $N(t)$, and finding the rates that correspond to the selected cell number time series in Figure \ref{fig:selfreg-3cases} \textbf{(C)} and \textbf{(D)}.
We couch our model in terms of density-dependent changes in birth and/or death rates (thus, population-number dependent, given a fixed total volume of the cell culture).
When the same net growth rate can arise from different density-dependent mechanisms, at the level of birth and death rates, the birth and death rates as functions of \emph{time} can appear markedly different.
For example, while in scenarios (I) and (II), the birth and death rates show monotonically increasing, sigmoidal shapes throughout time, in scenario (III), the birth rate has the shape of a concave-down quadratic function as shown by the ``$+$'' magenta curves in Figure \ref{fig:selfreg-3cases} \textbf{(E)} and \textbf{(F)}.\\\\
\subsection{Stage 2: Drug Efficacy} \label{sect:drug-effect}
\begin{figure}[H]
  \centering
  \includegraphics[scale=0.85]{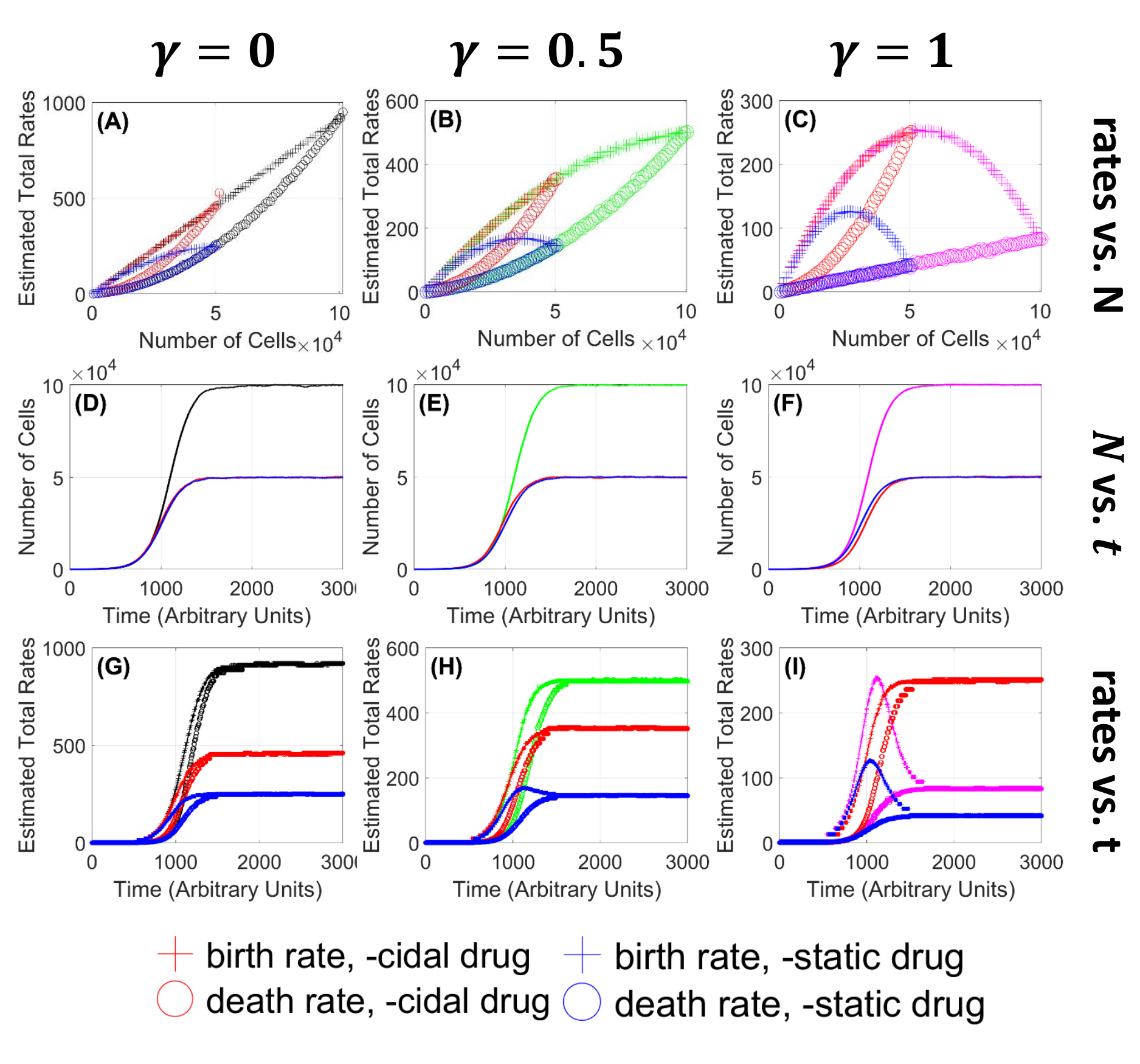}
  \caption{\textbf{Separating birth and death rates distinguishes the underlying -cidal versus -static action of drugs.}
  In each of the density-dependent cases (I), (II), (III), two different drugs reduce the cell population's carrying capacity to the same level (shown in red and blues curves in (D), (E), (F)), but the underlying mechanisms are different: increasing death rates (red curves) or decreasing birth rates (blue curves). 
  Black, green, and magenta curves represent scenarios (I), (II), and (III) without drugs. 
  Red curves represent the scenarios under a ``-cidal'' drug, and blue curves represent the scenarios under a ``-static'' drug. 
  Plus signs $(+)$ denote estimated birth rates; circles $(\circ)$ denote estimated death rates. \textbf{(A, D, G)}: scenario (I) with $\gamma=0$ (black curves), \textbf{(B, E, H)}: scenario (II) with $\gamma=0.5$ (green curves), \textbf{(C, F, I)}: scenario (III) with $\gamma=1.0$ (magenta curves). \textbf{(A, B, C)}: birth and death rates estimated from 100-trajectory ensembles. \textbf{(D, E, F)}: a representative trajectory without drug and two representative trajectories treated with drugs. 
  The red and curves trajectories have the same mean-field behavior but the drug mechanisms are different. \textbf{(G, H, I)}: estimated birth and death rates throughout time corresponding to the trajectories in \textbf{(D, E, F)}.}
  \label{fig:drugeffects_threescenarios}
\end{figure}

In this stage, the cell population is treated with a drug that cuts its carrying capacity in half, either by increasing the \textit{per capita} death rate $d_N$ or by decreasing the \textit{per capita} birth rate $b_N$.
If a drug acts by increasing the \textit{per capita} death rate, we refer to it as a drug with a ``-cidal" mechanism.
If a drug acts by lowering the \textit{per capita} birth rate, we refer to it as a drug with a ``-static" mechanism. 
If a drug combines both effects, we refer to such a treatment as having a mixed mechanism.
\\\\
Figure \ref{fig:drugeffects_threescenarios} shows our disambiguation results for the drug efficacy mechanisms for the three scenarios (I), (II), and (III) described in Section \ref{sect:auto-regulation}.
We simulate 100 trajectories of the cell population under each scenario with an initial population $N(t_0) = 10$, and the parameter values in Table \ref{table:simulation-parameters} under three drug efficacy cases: (i) without drug (black curves), (ii) with ``-cidal'' (death-promoting) drug (red curves), and (iii) with ``-static'' (birth-inhibiting) drug (blue curves).
Both of the drugs reduce the original carrying capacity $K$ to $K/2$. Under the ``-cidal'' drug, the \textit{per capita} birth and death rates are as follows
\begin{itemize}
    \item Scenario (I), -cidal: Drug increases the \textit{per capita} death rate, $(d_N)_{\text{cidal}}> (d_N)_{\text{original}}$: 
    \begin{align}
    (b_N)_{\text{cidal}} &= (b_N)_{\text{original}} = b_0,\\
    (d_N)_{\text{cidal}} &= d_0 + \dfrac{r}{(K/2)}N. 
    \end{align}
    The density dependence parameter $\gamma$ remains 0.
    \item Scenario (II), -cidal: Drug increases the \textit{per capita} death rate, $(d_N)_{\text{cidal}}> (d_N)_{\text{original}}$:
    \begin{align}
    (b_N)_{\text{cidal}} &= (b_N)_{\text{original}} = \max\left\{b_0 -0.5\dfrac{r}{K}N,0\right\} = \max\left\{b_0 -0.25\dfrac{r}{(K/2)}N,0\right\},\\
    (d_N)_{\text{cidal}} &= d_0 + 0.75\dfrac{r}{(K/2)}N = d_0 + 1.5\dfrac{r}{K}N. 
    \end{align}
    The density dependence parameter $\gamma$ changes from 0.5 to 0.25.
    \item Scenario (III), -cidal: Drug increases the \textit{per capita} death rate, $(d_N)_{\text{cidal}}> (d_N)_{\text{original}}$:
    \begin{align}
    (b_N)_{\text{cidal}} = (b_N)_{\text{original}} &= \max\left\{b_0 -\dfrac{r}{K}N,0\right\} =   \max\left\{b_0 -0.5\dfrac{r}{(K/2)}N,0\right\},\\
    (d_N)_{\text{cidal}} &= d_0 + 0.5\dfrac{r}{(K/2)}N = d_0 + \dfrac{r}{K}N.
    \end{align}
    The density dependence parameter $\gamma$ changes from 1 to 0.5.
\end{itemize}  
Under the ``-static'' drug, the \textit{per capita} birth and death rates are as follows
\begin{itemize}
    \item Scenario (I), -static: Drug decreases the \textit{per capita} birth rate, $(b_N)_{\text{static}} < (b_N)_{\text{original}}$:
    \begin{align}
    (d_N)_{\text{static}} &= (d_N)_{\text{original}} = d_0 + \dfrac{r}{K}N = d_0 + 0.5\dfrac{r}{(K/2)}N,\\ (b_N)_{\text{static}} &= \max\left\{b_0 - 0.5\dfrac{r}{K}N,0\right\}. 
    \end{align}
    The density dependence parameter $\gamma$ changes from 0 to 0.5.
    \item Scenario (II), -static: Drug decreases the \textit{per capita} birth rate, $(b_N)_{\text{static}} < (b_N)_{\text{original}}$:
    \begin{align}
    (d_N)_{\text{static}} &= (d_N)_{\text{original}} = d_0 + 0.5\dfrac{r}{K}N = d_0 + 0.25\dfrac{r}{(K/2)}N,\\
    (b_N)_{\text{static}} &= \max\left\{b_0 - 0.75\dfrac{r}{(K/2)}N\right\} = \max\left\{b_0 - 1.5\dfrac{r}{K}N\right\}. 
    \end{align}
    The density dependence parameter $\gamma$ changes from 0.5 to 0.75.
    \item Scenario (III), -static: Drug decreases the \textit{per capita} birth rate, $(b_N)_{\text{static}} < (b_N)_{\text{original}}$:
    \begin{align}
    (d_N)_{\text{static}} &= (d_N)_{\text{original}} = d_0,\\
    (b_N)_{\text{static}} &= \max\left\{b_0 - \dfrac{r}{(K/2)},0\right\}. 
    \end{align}
    The density dependence parameter $\gamma$ remains 1.
\end{itemize} 
\noindent
Figure \ref{fig:drugeffects_threescenarios} illustrates the effects of -cidal versus -static drugs in scenario (I) in the first column (panels \textbf{A, D, G}), scenario (II) in the second column (panels \textbf{B, E, H}), and scenario (III) in the third column (panels \textbf{C, F, I}).
As is evident in Figure \ref{fig:drugeffects_threescenarios} \textbf{(D, E, F)}, the observed cell number dynamics can be very similar in each scenario.  However, Figure \ref{fig:drugeffects_threescenarios} panels \textbf{(A, B, C)} and \textbf{(G, F, H)} show that the underlying birth and death processes that give rise to the dynamics can be very different. 
Specifically, in \textbf{(D, E, F)}, we see that the red and blue curves are almost indistinguishable. 
Thus, these scenarios could not easily be distinguished from the general shape of the growth curve alone.
However, to obtain the red curves, we keep the \textit{per capita} birth rates the same and increase the \textit{per capita} death rates, and to obtain the blue curves, it is the other way around--as illustrated in panels \textbf{(A, B, C)}.
The time-dependent birth and death rates in panels \textbf{(G, H, I)} also show significant differences.
In particular, the \textit{per capita} birth rates under the ``-static'' drug treatment (blue curves) are monotonically increasing in scenario (I) (density-dependent death rate, as shown in \textbf{(G)}), but show a pronounced increase and then decrease in scenario (III), as shown in \textbf{(I)}.
Thus, by extracting birth and death rates separately from cell number time series, we are able to disambiguate underlying drug mechanisms.
\subsection{Stage 3: Drug Resistance}\label{sect:drug-resistance}
\begin{figure}[H]
  \centering
  \includegraphics[scale=0.85]{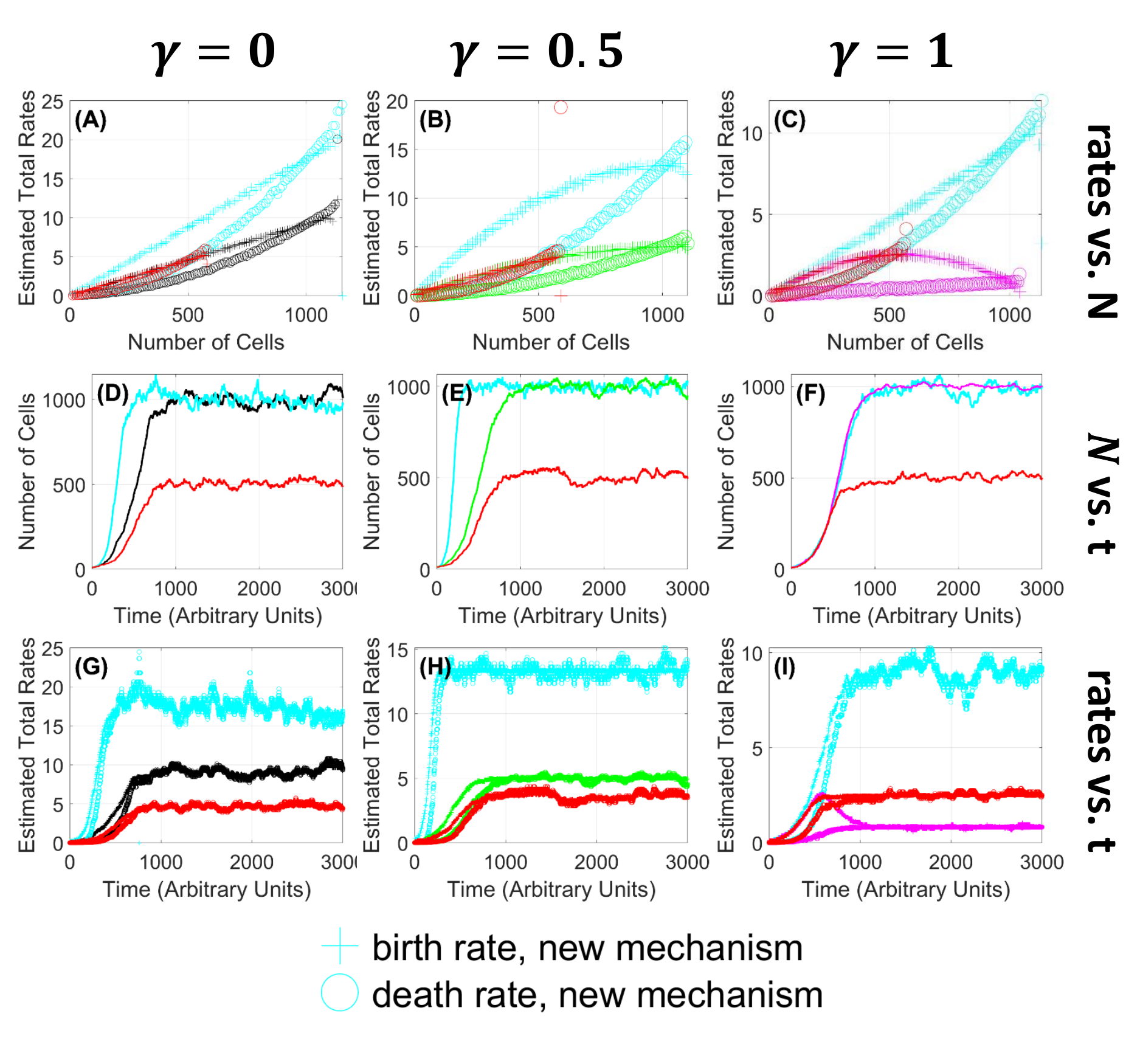}
  \caption{\textbf{Resolving separate birth vs.~death rates distinguishes different underlying mechanisms of resistance to -cidal drugs.}
  In each of the three density-dependent scenarios (I), (II), (III), a cell population can restore its carrying capacity after a ``-cidal'' drug treatment via different mechanisms: decreasing death rate to return to the original dynamics (shown in the black, green, magenta curves) or increasing birth rate (shown in the cyan curves). 
  \textbf{(A, B, C)}: estimated birth and death rates using an ensemble of 100 cell number trajectories. Plus signs $(+)$ denote estimated birth rates; circles $(\circ)$ denote estimated death rates. 
  \textbf{(D, E, F)}:  selected cell number trajectories. 
  \textbf{(G, H, I)}: estimated birth and death rates corresponding to the cell number trajectories in \textbf{(D, E, F)}. \textbf{(A, D, G)}: scenario (I) where the density dependence $\gamma = 0$. \textbf{(B, E, H)}: scenario (II) where the density dependence $\gamma = 0.5$. \textbf{(C, F, I)}: scenario (III) where the density dependence $\gamma = 1$. The red curves represent the case where the cell population has been treated with a ``-cidal'' drug. The black, green, and magenta curves represent the case where the cell population develops resistance by decreasing its \textit{per capita} death rate and returns to the original dynamics for the scenarios (I), (II), and (III) introduced in Section \ref{sect:auto-regulation}. The cyan curves represent the case in which the cell population develops resistance by increasing its \textit{per capita} birth rate.} 
  \label{fig:resistance-cidal}
\end{figure}

\begin{figure}[H]
  \centering
  \includegraphics[scale=0.85]{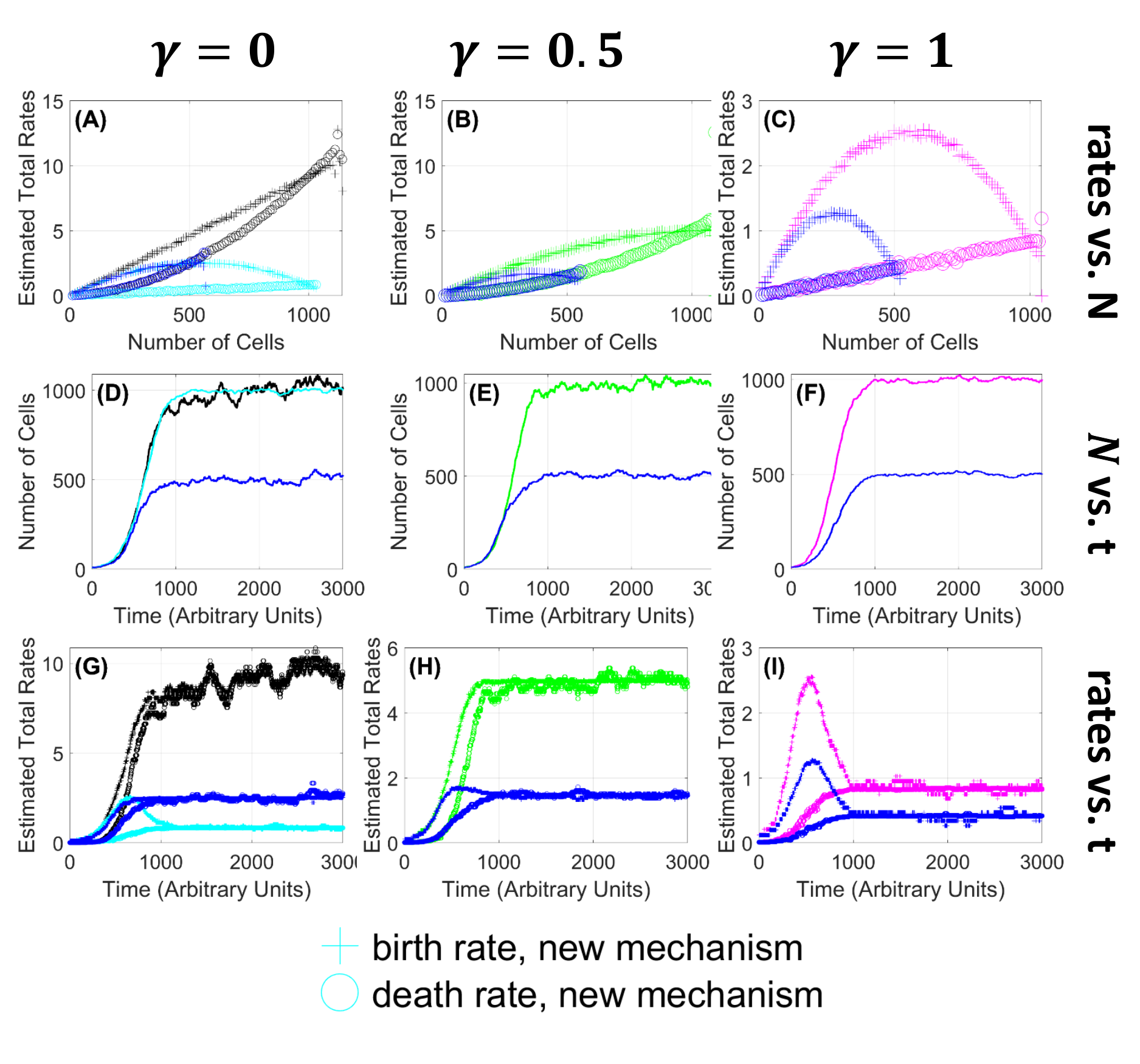}
  \caption{\textbf{Resolving separate birth vs.~death rates distinguishes different underlying mechanisms of resistance to -static drugs.}
  A cell population can restore its carrying capacity after a ``-static'' drug treatment via different mechanisms: increasing birth rate to return to the original dynamics (shown in the black, green, magenta curves) or decreasing death rate (shown in the cyan curves). The latter can happen only for scenario (I) where originally, the density dependence is fully in the death rate.
  \textbf{(A, B, C)}: estimated birth and death rates using an ensemble of 100 cell number trajectories. 
  Plus signs $(+)$ denote estimated birth rates; circles $(\circ)$ denote estimated death rates. \textbf{(D, E, F)}: selected cell number trajectories. \textbf{(G, H, I)}: estimated birth and death rates corresponding to the cell number trajectories in \textbf{(D, E, F)}:  selected cell number trajectories. \textbf{(A, D, G)}: scenario (I) where the density dependence $\gamma = 0$. \textbf{(B, E, H)}: scenario (II) where the density dependence $\gamma = 0.5$. \textbf{(C, F, I)}: scenario (III) where the density dependence $\gamma = 1$. 
  The blue curves represent the case where the cell population has been treated with a ``-static'' drug. The black, green, and magenta curves represent the case where the cell population develops resistance by decreasing its \textit{per capita} death rate and returns to the original dynamics for the scenarios (I), (II), and (III) introduced in Section \ref{sect:auto-regulation}. 
  The cyan curves represent the case in which the cell population develops resistance by increasing its \textit{per capita} birth rate.} 
  \label{fig:resistance-static}
\end{figure}

After having been treated with drugs that reduce their carrying capacities as described in Section \ref{sect:drug-effect}, cell populations can overcome the drug effects and revert to their original carrying capacities. 
We refer to this phenomenon as drug resistance. 
In this section, we demonstrate different mechanisms through which cell populations might develop drug resistance against a -cidal drug (Figure \ref{fig:resistance-cidal}) and against a -static drug  (Figure \ref{fig:resistance-static}),  for the three scenarios (I), (II), and (III) described in Section \ref{sect:auto-regulation}. 
In  simulating the scenarios for these two cases, we set the original carrying capacity to be $K = 10^3$, and keep the other original parameters to be the same as in Table \ref{table:simulation-parameters}. 
On this scale, the fluctuations are readily apparent in the traces; 
the method works robustly for larger values of $K$ as well.
Throughout, ``original'' means ``wild-type'' and ``before drug treatment''. 
We consider the case where the ``-cidal'' and ``-static'' drugs reduce the carrying capacity by a factor of 2.
\\\\
The effect of a drug and the cell population's resistance mechanism can be captured in part by a change in its carrying capacity, in part by a change in the distribution of density-dependent effects, described by $\gamma$, and in part by a change in the \textit{per capita} intrinsic/low-density birth and death rates, $b_0$ and $d_0$.
Figure \ref{fig:resistance-cidal} illustrates different mechanisms of drug resistance to the ``-cidal'' effect described in Section \ref{sect:drug-effect}.
\begin{itemize}
\item In scenario (I) as shown in Figure \ref{fig:resistance-cidal} \textbf{(A, B)}, the cell population can develop resistance either by decreasing its \textit{per capita} death rate back to the original rate:
\begin{align}
(d_N)_{\text{cidal, resistant}} &= (d_N)_{\text{original}} = d_0 + \dfrac{r}{2(K/2)}N,\\
(b_N)_{\text{cidal, resistant}} &= (b_N)_{\text{cidal}} = (b_N)_{\text{original}} = b_0,
\end{align}
or by increasing its \textit{per capita} intrinsic birth rate to $b_0 + r = 2r + d_0$:
\begin{align}
(b_N)_{\text{cidal, resistant}} &= b_0 + r = 2r + d_0,\\
(d_N)_{\text{cidal, resistant}} &= (d_N)_{\text{cidal}} = d_0 + \dfrac{r}{(K/2)}N = d_0 + \dfrac{2r}{K}N,
\end{align}
which leads to the \textit{per capita} intrinsic net growth rate $r$ increasing to $2r$. Such an increase in the intrinsic cell division rate could potentially arise through mutation. 
(Why such a mutation would not already have been exploited in the wild-type cell line is a question beyond the scope of this paper.)
In both drug resistance scenarios, the density dependence parameter $\gamma$ remains 0, which suggests no significant change in the cell-to-cell interaction modality.
\item In scenario (II) as shown in Figure \ref{fig:resistance-cidal} \textbf{(C, D)}, the cell population can develop resistance by either decreasing its \textit{per capita} death rate back to the original rate:
\begin{align}
(d_N)_{\text{cidal, resistant}} &= (d_N)_{\text{original}} = d_0 + 0.5\dfrac{r}{2(K/2)}N,\\
(b_N)_{\text{cidal, resistant}} &= (b_N)_{\text{cidal}} = (b_N)_{\text{original}} = \max\left\{b_0 - 0.5\dfrac{r}{K}N,0\right\},
\end{align}
or by increasing its \textit{per capita} intrinsic birth rate to $b_0 + 2r = 3r + d_0$:
\begin{align}
(b_N)_{\text{cidal, resistant}} &= \max\left\{3r + d_0 -0.5\dfrac{3r}{K}N,0\right\},\\
(d_N)_{\text{cidal, resistant}} &= (d_N)_{\text{cidal}} = d_0 + 1.5\dfrac{r}{2(K/2)}N = d_0 + 0.5\dfrac{3r}{K}N,
\end{align}
which shows that the \textit{per capita} intrinsic net growth rate $r$ would have to increase to $3r$. 
Such an increase in the intrinsic cell division rate could potentially arise through mutation.
In both drug resistance scenarios, the density dependence parameter $\gamma$ changes from 0.25 back to 0.5, which suggests a change in cell-to-cell interaction modality. 
Note that the drug resistance mechanism through death in this scenario is different from scenario (I), because in scenario (I), the \textit{per capita} death rate decreases only due to increased carrying capacity, while in this scenario, the \textit{per capita} death rate decreases also due to decreased density dependence of death (i.e. $(1-\gamma)$ changes from 0.75 to 0.5). 
\item In scenario (III) as shown in Figure \ref{fig:resistance-cidal} \textbf{(E, F)}, the cell population can become drug resistant either by decreasing its \textit{per capita} death rate back to the original rate:
\begin{align}
(d_N)_{\text{cidal, resistant}} &= (d_N)_{\text{original}} = d_0, \\
(b_N)_{\text{cidal, resistant}} &= (b_N)_{\text{cidal}} = (b_N)_{\text{original}} = \max \left\{b_0 - \dfrac{r}{K}N,0\right\},
\end{align}
or by increasing its \textit{per capita} birth rate to \begin{align}
(b_N)_{\text{cidal, resistant}} &= b_0,\\
(d_N)_{\text{cidal, resistant}} &= (d_N)_{\text{cidal}} = d_0 + \dfrac{r}{2(K/2)}N.
\end{align}
In the first scenario (drug resistance mechanism via modified death rate), the density dependence parameter $\gamma$ changes from 0.5 back to 1, while in the drug resistance mechanism through birth, the density dependence parameter $\gamma$ changes from 0.5 to 0. Both of these scenarios suggest changes in the cell-to-cell interaction modalities. The latter suggests a significant change from full density dependence in birth (before drug treatment) to full density dependence in death (after ``-cidal'' drug treatment and resistance). Note that the \textit{per capita} intrinsic rates, $b_0$ and $d_0$, remain the same. 
\end{itemize}
Figure \ref{fig:resistance-cidal} shows that having been treated with a ``-cidal'' drug, the cell population can develop resistance either by reverting to its original dynamics--the red curves change back to the black, green, and magenta curves for scenarios (I), (II), and (III) respectively in the figure, or by increasing its \textit{per capita} birth rate as illustrated by the cyan curves. 
We may call the latter drug resistance mechanism ``enhanced fecundity''  or ``hyper-birth.'' 
Without computing the birth and death rates explicitly, we observe from cell number time series that if the resistant cell population (cyan curves) reaches its original carrying capacity earlier than the wild-type population (black, green, magenta curves) as in Figure \ref{fig:resistance-cidal} \textbf{(B, D)} or if the typical fluctuations around the mean population size are visibly larger than the fluctuations of the wild-type as in Figure \ref{fig:resistance-cidal} \textbf{(F)}, we  
may hypothesize that
the population has developed drug resistance through the ``hyper-birth'' mechanism. 
\newpage
\noindent
Figure \ref{fig:resistance-static} illustrates different mechanisms of drug resistance to the ``-static'' effect described in Section \ref{sect:drug-effect}.
\begin{itemize}
\item In scenario (I) as shown in Figure \ref{fig:resistance-static} \textbf{(A, B)}, the cell population can become drug resistant either by increasing its \textit{per capita} birth rate back to the original rate:
\begin{align}
(b_N)_{\text{static, resistant}} &= (b_N)_{\text{original}} = b_0,\\
(d_N)_{\text{static, resistant}} &= (d_N)_{\text{static}} = (d_N)_{\text{original}} = d_0 + \dfrac{r}{K}N,
\end{align}
or by decreasing its \textit{per capita} death rate:
\begin{align}
(d_N)_{\text{static, resistant}} &= d_0,\\
(b_N)_{\text{static, resistant}} &= (b_N)_{\text{static}} = b_0 - 0.5\dfrac{r}{(K/2)}N = \max\left\{b_0 - \dfrac{r}{K}N, 0\right\}.
\end{align}
In the drug resistance mechanism through birth, the density dependence parameter $\gamma$ changes from 0.5 back to 0, while in the drug resistance mechanism through death, the density dependence parameter $\gamma$ changes from 0.5 to 1. Both of these scenarios suggest changes in the cell-to-cell interaction modalities. The latter suggests a significant change from full density dependence in death (before drug treatment) to full density dependence in birth (after ``-static'' drug treatment and resistance).
\item In scenario (II) as shown in Figure \ref{fig:resistance-static} \textbf{(C, D)}, the cell population can develop resistance by either increasing its \textit{per capita} birth rate back to the original rate:
\begin{align}
(b_N)_{\text{static, resistant}} &= (b_N)_{\text{original}} = b_0 - 0.5\dfrac{r}{K}N,\\
(d_N)_{\text{static, resistant}} &= (d_N)_{\text{static}} = (d_N)_{\text{original}} = d_0 + 0.25\dfrac{r}{(K/2)}N = d_0 + 0.5\dfrac{r}{K}N,
\end{align}
or by decreasing its \textit{per capita} intrinsic death rate to $d_0 - 2r$:
\begin{align}
(d_N)_{\text{static, resistant}} &= d_0 - 2r + 0.5\dfrac{3r}{K}N \label{eq:neg-death-caseII},\\
(b_N)_{\text{static, resistant}} &= (b_N)_{\text{static}} = \max\left\{b_0 - 0.5\dfrac{3r}{K}N,0\right\}.
\end{align}
In the drug resistance mechanism through birth, the density dependence parameter $\gamma$ changes from 0.75 back to 0.5, which suggests a change in the cell-to-cell interaction modality. In the drug resistance mechanism through death, the new \textit{per capita} intrinsic death rate, $d_0 - 2r$, can be negative, which is not biologically meaningful. 
\item In scenario (III) as shown in Figure \ref{fig:resistance-static} \textbf{(E, F)}, the cell population can develop resistance by either increasing its \textit{per capita} birth rate back to the original rate:
\begin{align}
(b_N)_{\text{static, resistant}} &= (b_N)_{\text{original}} = \max\left\{b_0 - \dfrac{r}{K}N\right\},\\
(d_N)_{\text{static, resistant}} &= (d_N)_{\text{static}} = (d_N)_{\text{original}} = d_0,
\end{align}
or by decreasing its \textit{per capita} intrinsic death rate to $d_0 - r$:
\begin{align}
(d_N)_{\text{static, resistant}} &= d_0 - r \label{eq:neg-death-caseIII},\\
(b_N)_{\text{static, resistant}} &= (b_N)_{\text{static}} = \max\left\{b_0 - \dfrac{2r}{K}N, 0\right\}.
\end{align}
In the drug resistance mechanism through birth, the density dependence parameter $\gamma$ remains 1, which suggests no significant change in the cell-to-cell interaction modality. In the drug resistance mechanism through death, the new \textit{per capita} intrinsic death rate, $d_0 - r$, can be negative, which is not biologically meaningful. 
\end{itemize}
\noindent
Figure \ref{fig:resistance-static} shows that having been treated with a ``-static'' drug, the cell population can develop resistance either by reverting to its original dynamics--the blue curves change back to the black, green, and magenta curves for scenarios (I), (II), and (III) respectively in the figure--or by decreasing its \textit{per capita} death rate as illustrated by the cyan curves. 
We may call the latter drug resistance mechanism ``reduced mortality" or ``hypo-death.'' 
We note that the decreased \textit{per capita} death rate can become algebraically negative and not biologically meaningful, as seen in Equations \eqref{eq:neg-death-caseII} and \eqref{eq:neg-death-caseIII}, which is consistent with the fact that drug resistance has previously been considered mainly for ``-cidal'' drugs, not ``-static'' drugs, in the literature, cf.~\parencite{brauner2016distinguishing}.
However, in contrast to some recent literature \parencite{brauner2016distinguishing}, in this paper, we propose the possibility of  mechanisms through which cell populations can overcome the ``-static'' effect (birth inhibition) of drugs--that is, increasing the \textit{per capita} birth rates back to the original rates, as seen in the black, green, and magenta curves in Figure \ref{fig:resistance-static}. 
For instance if, through preexisting genetic variation, the cell population contained a mutant with an alternative sequence for the protein by which the drug targets the cell, then as this variant propagated in favor of the principal variant, the cell line could develop resistance to the ``-static'' drug.
It is interesting to observe in Figure \ref{fig:resistance-static} \textbf{(G)} that even after being with a ``-static'' drug that inhibits birth, the cell population can develop resistance by reducing birth rates \textit{throughout time}--as we can see the cyan curves are lower than the blue curves as time increases.
\\\\
We note that for scenario (I) where $\gamma = 0$, we observe a second possible drug resistance mechanism, in which the cell population decreases its \textit{per capita} death rate without making it negative. In this scenario, the cell population also changes its density dependence parameter from $\gamma = 0$ to $\gamma = 1$ as it becomes resistant to the ``-static'' drug.

\section{Likelihood-Based Inference}\label{sect:likelihood}
\begin{figure}[H]
  \centering
  \includegraphics[scale=0.85]{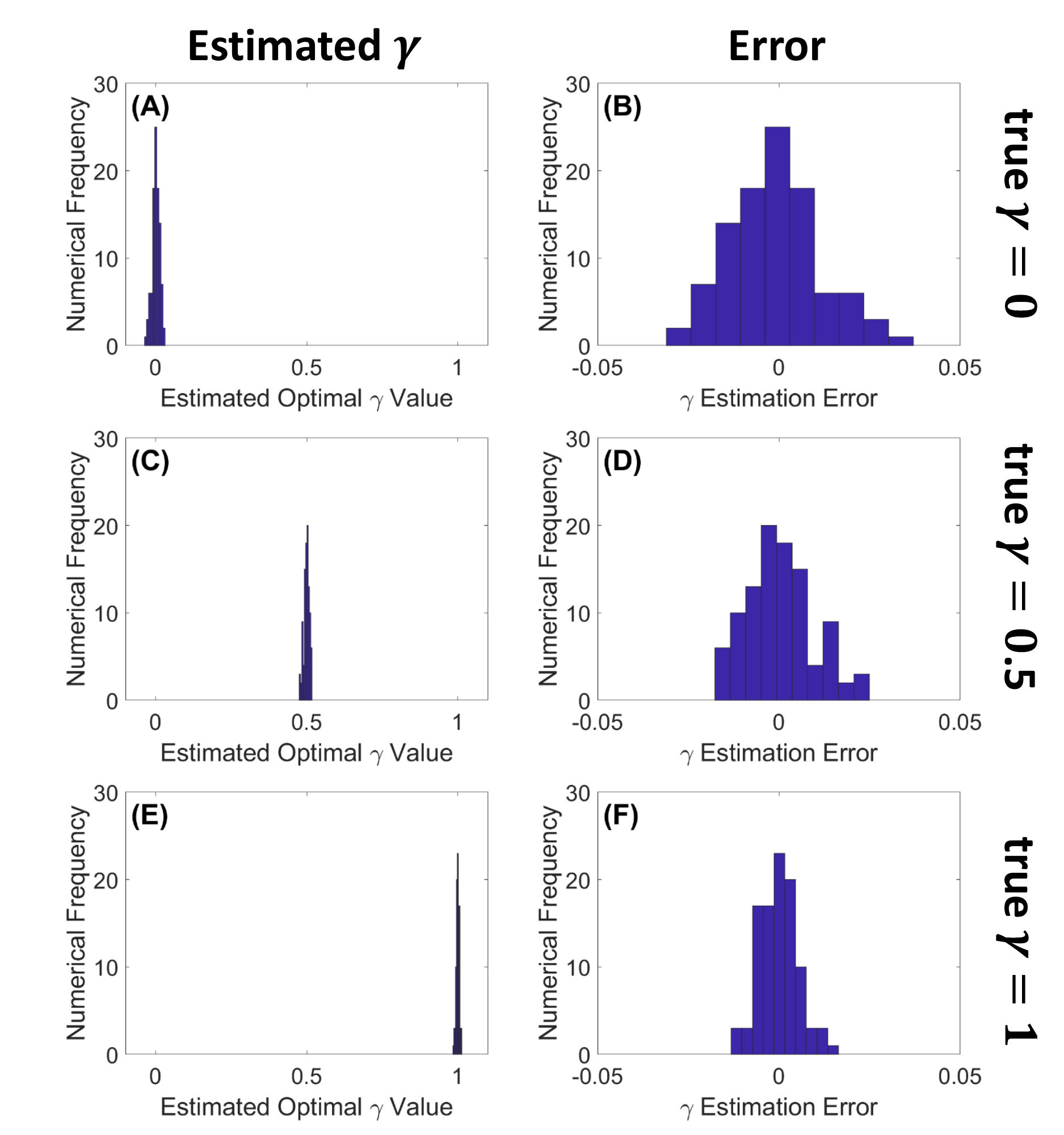}
  \caption{\textbf{Numerical solutions to the optimization problem for different density dependence scenarios are clearly separated.} We plot empirical distributions of estimated $\gamma$ values, and corresponding errors, for an ensemble of individual cell number time series.
  \textbf{(A, B)}: cell number time series simulated with $\gamma = 0$; \textbf{(C, D)}: cell number time series simulated with $\gamma = 0.5$; \textbf{(E, F)}: cell number time series simulated with $\gamma = 1$. 
  \textbf{(A, C, E)}: distributions of estimated $\gamma$; \textbf{(B, D, F)}: distributions of the corresponding estimation errors, defined as $(\gamma_{\text{data}} - \gamma_{\text{estimated}})$. 
  We observe that the distributions of the estimated $\gamma$ have small variances, and the errors are approximately normally distributed.
  Here we used time series of length $T=90,000$ points (timestep $\Delta t=1/30,$ total time $3000$).}
  \label{fig:likelihood-distributions}
\end{figure}

In the instance that there is only a single cell number trajectory, we would like to be able to assert how likely it is that the time series belongs to one of several scenarios, parameterized by the density dependence parameter $\gamma$. 
This question leads us to consider a maximum likelihood approach. 
\\\\
Let a cell number time series ${\bm{X}}_{\text{data}} = [x_0, x_1, \ldots, x_T]$ be a realization for the normally distributed random variable $\bm{X} = [X(t_0), X(t_1), \ldots, X(t_T)]$, which approximates the discrete random variable $\bm{N} = [N(t_0), N(t_1), \ldots, N(t_T)]$ as discussed in Section \ref{sect:data-simulation}. 
For clarity, we denote $x_j$ as $x_{j,\text{data}}, j = 0,\ldots,T$.
Recall that $X(t)$ follow a Gaussian birth-death process\footnote{This approximation requires sufficiently large summed birth and death rates.} characterized by the parameter set $\Theta_{\text{data}} = \{b_{0,\text{data}}, d_{0,\text{data}}, \gamma_{\text{data}}, K_{\text{data}}\}$. These parameters determine the birth and death rates of the birth-death process from which the time series is generated. In particular, the \textit{per capita} birth and death rates are defined as follows:
\begin{align}
    b_{X_{j,\text{data}}} &= \max\left\{b_{0,\text{data}} - \gamma_{\text{data}} \dfrac{r_{\text{data}}}{K_\text{data}}X_{j,\text{data}},\,0\right\},\\
    d_{X_{j,\text{data}}} &= d_{0,\text{data}} + (1-\gamma_{\text{data}})\dfrac{r_{\text{data}}}{K_{\text{data}}}X_{j,\text{data}},
\end{align}
where $r_{\text{data}} = b_{0,\text{data}} - d_{0,\text{data}}$. To test whether a given time series $\bm{X}_{\text{data}}$ belongs to a scenario characterized by the parameter set $\Theta_{\text{test}} = \{b_{0,\text{test}}, d_{0,\text{test}}, \gamma_{\text{test}}, K_{\text{test}}\}$, we evaluate the log likelihood function at the time series:
\begin{align}
    &\mathcal{L}(\bm{X}_{\text{data}}|\Theta_{\text{test}}) = \underbrace{\ln\Big(P(x_{0,\text{data}}) \Big)}_{=\ln(1)}+\sum_{j=1}^{T-1}\ln\Big(P(x_{j+1,\text{data}}|x_{j,\text{data}}, \Theta_{\text{test}}) \Big)\\
&= \sum_{j=1}^{T-1}\dfrac{1}{2}\ln\left(\dfrac{1}{2\pi x_{j,\text{data}}(b_{\text{test},x_{j,\text{data}}}+d_{\text{test},x_{j,\text{data}}})\Delta t}\right) - \dfrac{1}{2}\dfrac{\Big(x_{j+1,\text{data}} - x_{j,\text{data}}-x_{j,\text{data}}(b_{\text{test},x_{j,\text{data}}}-d_{\text{test},x_{j,\text{data}}})\Delta t\Big)^2}{x_{j,\text{data}}(b_{\text{test},x_{j,\text{data}}}+d_{\text{test},x_{j,\text{data}}})\Delta t}, \label{eq:log-likelihood}
\end{align}
where
\begin{align}
    b_{\text{test},x_{j,\text{data}}} &= \max\left\{b_{0,\text{test}} - \gamma_{\text{test}} \dfrac{r_{\text{test}}}{K_\text{test}}x_{j,\text{data}},\,0\right\},\\
    d_{\text{test},x_{j,\text{data}}} &= d_{0,\text{test}} + (1-\gamma_{\text{test}})\dfrac{r_{\text{test}}}{K_{\text{test}}}x_{j,\text{data}},\\
    r_{\text{test}} &= b_{0,\text{test}} - d_{0,\text{test}}.
\end{align}
\noindent
Suppose we know $b_{0,\text{data}}, d_{0,\text{data}}$, and $K_{\text{data}}$. That is, suppose that $b_{0,\text{test}} = b_{0,\text{data}}$, $d_{0,\text{test}} = d_{0,\text{data}}$, and $K_{\text{test}} = K_{\text{data}}$. Given one cell number time series, to infer which density dependence scenario the data mostly likely belongs, we treat the log-likelihood function as a function $f$ of $\gamma:= \gamma_{\text{test}}$, and find $\gamma \in [0,1]$ that maximizes $f(\gamma)$. 
We thus formulate a constrained nonlinear optimization problem as follows:
\begin{align}
    \displaystyle{\max_{\gamma }} f(\gamma) = \mathcal{L}(\bm{X}_{\text{data}}|\Theta_{\text{test}}) \quad \text{subject to} \quad 0 \leq \gamma \leq 1. \label{eq:max-over-gamma}
\end{align}
For shorter notation, here we denote $b_{0,\text{test}}$ and $b_{0,\text{data}}$ as $b_0$, $d_{0,\text{test}}$ and $d_{0,\text{data}}$ as $d_0$, $K_{\text{test}}$ and $K_{\text{data}}$ as $K$, and $x_{j,\text{data}}$ as $x_j$.  We calculate the first derivative $df/d\gamma$ in Appendix \ref{appendix:appx-log-likelihood} and find critical points by solving $\dfrac{df}{d\gamma} = 0, \gamma \in [0,1]$ numerically using the Bisection method on the interval $[0-\Delta \gamma, 1+ \Delta \gamma], \Delta \gamma = 0.5 > 0$. 
For a discussion on the maximality of the critical points, 
please refer to Appendix \ref{appendix:appx-log-likelihood}.\\\\
Given multiple samples of cell number time series (e.g.~from multiple experimental trials), we obtain an empirical distribution of solutions $\gamma$ to the optimization problem \eqref{eq:max-over-gamma}. 
In Figure \ref{fig:likelihood-distributions}, for each of the three scenarios (I) $\gamma_{\text{data}} = 0$, (II) $\gamma_{\text{data}} = 0.5$, and (III) $\gamma_{\text{data}} = 1$, we plot the results upon solving the optimization problem 100 times for 100 independent time series, and obtain a distribution of estimated $\gamma$ parameters.  
In addition we obtain a distribution of the estimation error, defined as the absolute difference $(\gamma_{\text{data}} - \gamma_{\text{estimated}})$, where $\gamma_{\text{estimated}}$ is the numerical solution to the optimization problem \eqref{eq:max-over-gamma}. 
The values of the parameters $b_0$, $d_0$, and $K$ used in data simulation are the same as in Table \ref{table:simulation-parameters}. 
The empirical means and variances of the estimated $\gamma$ values and estimation errors for the three scenarios (I), (II), and (III) are listed in Table \ref{table:maxgamma-estimated}.
\begin{table}[H]
\caption{Numerical solution to the optimization problem \eqref{eq:max-over-gamma} for a time series of length $T=90,000$ points (timestep $\Delta t=1/30,$ total time $3000$ arbitrary units).} 
\centering 
\begin{tabular}{c c c c c} 
\hline\hline 
True $\gamma$ Value & Mean Estimated $\gamma$ & Variance of Estimated $\gamma$ & Mean Error & Error Variance \\ [0.5ex] 
\hline\hline
0 & 0.0010 & $1.4858 \times 10^{-4}$ & $-0.0010$ & $1.4858 \times 10^{-4}$\\ 
0.5 & 0.4996 & $8.4605 \times 10^{-5}$ & $2.7218 \times 10^{-5}$ & $8.4605 \times 10^{-5}$\\
1 & 1.0002 & $3.6266 \times 10^{-4}$ & $-1.5752 \times 10^{-4}$
 & $2.7218 \times 10^{-5}$ \\
\hline 
\end{tabular}
\label{table:maxgamma-estimated} 
\end{table}
\noindent
We note that the mean values of $\gamma$ for the three scenarios (I), (II), and (III) are separated by margins that are an order of magnitude larger than the standard errors of the estimates.  Thus, for the data generated by our birth/death simulations, the distribution  the density-dependent effects can clearly be distinguished in terms of fully a birth-rate effect, fully a death-rate effect, or an evenly mixed effect.  
\noindent

\section{Conclusion and Discussion} \label{sect: discussion}
In order to infer density-dependent population dynamics mechanisms from data, we separately identify density-dependent \textit{per capita} birth and death rates from net growth rates using the method described in Section \ref{sect:direct-estimation} and infer whether density dependence is manifest in the birth process, death process, or some combination of the two. 
Our method involves directly estimating the mean and variance of cell number increments, as functions of population size, and expressing birth and death rates in terms of these two statistics. 
In order to obtain the mean and variance with tolerable accuracy, we compute them from an ensemble of cell number time series (e.g. multiple experiments). 
We analyze the accuracy of this method and
derive analytical expressions for the theoretical expected errors and variance of errors in estimating birth and death rates as functions of the bin size (details are in Appendix \ref{appendix:appx-error-analysis-direct-method}).
We discover that small bin sizes do not necessarily result in small errors in estimating birth and death rates, due to small sample sizes. 
In fact, we find that intermediate bin sizes are optimal.
Our error analysis also shows that if the intrinsic \textit{per capita} net growth rate $r$ is large relative to the carrying capacity $K$, then the expected error in estimating the mean cell number increment is high, as shown in Equation \eqref{eq:error-mean-expected}, which suggests that the estimation is not as good for fast-producing cell types.
\\\\
Our method is distinct from other methods in the literature.  
It provides a novel perspective on the problem of stochastic parameter identification.
Existing methods typically require numerical solution of a high-dimensional optimization problem, e.g.~in a Bayesian inverse problem setting \parencite{calvetti2007introduction} or a likelihood function maximization framework. \parencite{Crawford:2014} constructs an expectation-maximization algorithm to identify birth and death rates for general birth-death processes. 
This method enjoys fast convergence and benefits from
an elegant formulation of conditional expectations in terms of convolutions of transition probabilities.
Their approach results from solving a maximum likelihood problem.   
In contrast, we suggest a simple direct estimation approach that accurately extracts birth and death rates from the conditional first and second moments of the cell number time series data. 
Our method focuses on fundamental principles of stochastic processes and utilizes the nonzero variance of cell number increments.
Aside from \parencite{Crawford:2014}, to the best of our knowledge, other work addressing disambiguation of birth and death rates has been confined to \emph{linear} birth-death processes.  For example, \parencite{Liu:2018} uses a Bayesian approach to parameter estimation for linear birth-death models in order to quantify the effects of changing drug concentrations.
Here, we also consider different drug treatment scenarios,  but in the context of nonlinear, logistic population models rather than linear growth models.  
\parencite{Ferlic:2019, roney2020estipop} estimates birth and death rates as functions of time for a continuous-time branching process.  Their method applies to multi-type cell populations and 
is illustrated with density-independent \textit{per capita} birth and death rates.
In contrast, our framework encompasses density-dependent \textit{per capita} rates.
\\\\
Our direct estimation method is a data-hungry approach.  
As an alternative, for small sample sizes, we also present a  maximum likelihood approach, in which we evaluate the log-likelihood function and maximize it over the density dependence parameter $\gamma\in[0,1]$. This approach, which involves solving a one-dimensional constrained nonlinear optimization problem, is limited to the assumption that the other system parameters are known. 
\\\\
The significance of both approaches is the application in studying treatments of pathogens and their resistance to the treatments. Specifically, in Section \ref{sect:applications}, we consider the
scenario where a homogeneous cell population goes through three stages: (1)
grows naturally to its carrying capacity, (2) is treated with a drug that
reduces its carrying capacity, and (3) overcomes the drug effect to gain back
its carrying capacity. Our method allows us to identify whether each stage
happens through the birth process, death process, or some combination of the two.
Our analysis contributes to disambiguating underlying mechanisms such as exploitation vs.~interference competition in ecology, bacteriostatic vs.~bactericidal antibiotics in clinical treatments, and enhanced fecundity vs.~reduced mortality in pathogens' defense against drug treatments, which we may define as drug resistance.
The mechanisms shown in this paper can help explain biological phenomena 
and may suggest novel approaches for engineering synthetic biological systems.
More microscopic mechanisms within the birth process or death process, such as  inactivating mutations of the gene for p53 protein \parencite{baguley2010multiple}, are beyond the scope of the model in this paper. 
\\\\
In Section \ref{sect:drug-effect},  we show how to apply our method to distinguish the action of  ``-static'' (birth-inhibiting) versus ``-cidal'' (death-promoting) drugs. 
However, the classification of drugs as being ``-static'' or ``-cidal'' is complicated by potentially stochastic factors such as external growth conditions 
\parencite{Pankey:2004}. 
For bacterial infections in a clinical setting, the ``-static/-cidal'' distinction is defined in terms of drug concentrations--specifically in terms of the ratio between Minimum Inhibitory Concentration (MIC) and Minimum Bactericidal Concentration (MBC).
The Minimum Inhibitory Concentration (MIC) is defined as the lowest drug concentration that prevents visible growth,
and the Minimum Bactericidal Concentration (MBC) is defined as the lowest drug concentration that results in a $99.99\%$ decrease in the initial population size over a fixed period of time.
Bacteriostatic drugs have been defined as those for which the ratio of the MBC to the 
MIC is larger than 4. 
Bactericidal drugs are those for which the ratio is $\leq 4$ \parencite{WaldDickler:2018}. 
Including the differential effects of drugs at larger or smaller concentrations will be an interesting direction for expanding our birth/death rate analysis in future work.\\\\
In Section \ref{sect:drug-resistance},  we use our direct estimation method to disambiguate different drug resistance mechanisms. 
In our paper, we define ``drug resistance" as the cell population's ability to overcome the drug effect and gain back its original carrying capacity.
However, the term ``drug resistance" is used to  mean different things in the research literature. 
For example, in Davison et al.~2000 \parencite{davison2000antibiotic}, drug resistance is defined in terms of the drug concentration needed to inhibit growth or kill the pathogen. 
Brauner et al.~2016 \parencite{brauner2016distinguishing} quantify cell populations overcoming drug effects in terms of MIC and the minimum time needed to kill the pathogens (MDK). Based on these two measures, MIC and MDK, the pathogens' defense against the drug can be called drug tolerance, persistence, or resistance. For future work, we will look into different definitions of ``drug resistance''.  
\\\\
For the present study, we confine our investigation to simulated data because of several factors.  
First, generating large ensembles of cell population trajectories is expensive, although high-throughput methods continue to accelerate the pace of data generation \parencite{gopalakrishnan2020low}.  
In a typical bioreactor, the data available are optical density time series, rather than direct cell number measurements.
In theory, the relation between optical density and cell count is expected to be linear. 
Unfortunately, that is not always the case. 
McClure et al.~1993 \parencite{McClure:1993} show that it can be second order and Stephens et al.~1997 \parencite{Stephens:1997} show that it can be third order. 
Moreover, Stevenson et al.~2016 \parencite{Stevenson:2016} show that the relation between cell count and optical density varies for different cell sizes and shapes, as well as other properties such as the index of refraction of the media. 
Some experimental calibration techniques have been developed to overcome these discrepancies, such as Francois et al.~2005 \parencite{Francois:2005} and Beal et al.~2020 \parencite{Beal:2020}. 
Finally, experimental data may include measurement noise that obscures finite population driven density fluctuations. Swain et al.~2016 \parencite{Swain:2016} attempts to estimate net growth rates from optical density data using a Gaussian process framework. In contrast, we would like disambiguate net growth rate into separate birth and death rates.
Extending our method to take into account the mapping from cell number to noisy optical density measurements is an interesting subject for future work.\\\\
As mentioned in the Introduction (Section \ref{sect:introduction}), throughout the paper, we interpret the density dependence term (interaction between individuals) as competition, which either reduces birth rates or increases death rates. 
However, in some situations, interactions among individuals can be cooperative, and increase the birth rate or reduce death rate with increasing population size \parencite{bhowmick2015cooperation}. 
To address this possibility, in future work one might introduce to a cooperation parameter $c \geq 0$:
\begin{align}
    \dfrac{d\phi}{dt} = r\phi - \dfrac{r}{K}\phi^2 &= r\phi + \underbrace{c\dfrac{r}{K}\phi^2}_{\text{cooperation}} - \underbrace{(1+c)\dfrac{r}{K}\phi^2}_{\text{competition}}.
\end{align}
One may interpret the cooperation term $c\dfrac{r}{K}\phi^2$ as a positive interaction between individuals that increases cell population growth. 
One could parameterize this term with parameter $\gamma_c$, to quantify how much of the cooperation increases birth and how much of the cooperation decreases death. 
Similarly, one may interpret the competition term $(1+c)\dfrac{r}{K}\phi^2$ as a negative interaction between individuals that reduces cell population growth. 
One could parameterize the competition term with parameter $\gamma_{\sim c}$ to quantify how much of the competition decreases birth and how much of the competition increases death:
\begin{align}
    \dfrac{d\phi}{dt} =& r\phi + \underbrace{\gamma_c c \dfrac{r}{K}\phi^2}_{\text{cooperation}} + \underbrace{(1-\gamma_c)c\dfrac{r}{K}\phi^2}_{\text{cooperation}} - \underbrace{\gamma_{\sim c}(1+c)\dfrac{r}{K}\phi^2}_{\text{competition}} - \underbrace{(1-\gamma_{\sim c})(1+c)\dfrac{r}{K}\phi^2}_{\text{competition}}\\
    =& \underbrace{\Big(b_0\phi + \gamma_cc\dfrac{r}{K}\phi^2 - \gamma_{\sim c}(1+c)\dfrac{r}{K}\phi^2\Big)}_{\text{birth}}\\
    &- \underbrace{\Big(d_0\phi - (1-\gamma_c)c\dfrac{r}{K}\phi^2 + (1-\gamma_{\sim c})(1+c)\dfrac{r}{K}\phi^2\Big)}_{\text{death}}.
\end{align}
This study would provide a new perspective on modeling and analyzing the Allee effect and help disentangle positive and negative density dependence.
Exploring these and other extensions provide interesting directions for future investigation.
\newpage
\printbibliography
\newpage
\appendix 
\begin{table}[H]
\section{Model Parameters Used in Simulation}
\caption{Model Parameters Used in Simulation} 
\centering 
\begin{tabular}{c c c} 
\hline\hline 
Parameter & Value & Unit \\ 
\hline 
$b_0$ & 1.1/120 & 1/time  \\ 
$d_0$ & 0.1/120  & 1/time \\
$r$ & 1/120 & 1/time  \\
$K$ & $10^5$ & Dimensionless\\ 
\hline 
\end{tabular}
\label{table:simulation-parameters} 
\end{table}
\section{Error Analysis of the Direct Estimation Method} \label{appendix:appx-error-analysis-direct-method}
As described in Section \ref{sect: direct-method}, we discretize all the values of cell number across the whole ensemble of trajectories into bins. 
Denote the bin size as $\eta$.
The left end point $N_k$ of the $k$th bin $[N_k, N_k + \eta)$ with $k = 1, 2, \ldots, k_{\max}$ is equal to $N_k := N_{\min} + (k-1)\eta$, where $N_{\min}$ is the smallest value of cell number across the whole ensemble of trajectories.
In many instances, $N_{\min}=N(t_0)$, the initial population size.
The total number of bins $k_{\max} \in \mathbb{Z}^+$ is equal to $\Big\lceil\dfrac{N_{\max} - N_{\min}}{\eta} \Big\rceil$, where $N_{\max}$ is the largest value of cell number across the whole ensemble of trajectories, and $\lceil n \rceil$ is the smallest integer not less than $n$.
The $i$th cell number element to have landed in the $k$th bin $[N_k,N_k + \eta)$ is equal to $N_k + \eta_i$.
For simplicity, we make the approximation that for each bin, the random variables $\eta_i$ are i.i.d.~and uniformly distributed on $[0,\eta)$.
We expect  this approximation to be reasonably accurate when the bin size $\eta$ is small  enough that a given trajectory is unlikely to land in any particular bin twice in succession; the approximation may become inaccurate for excessively large bin sizes.
In light of this uniform distribution assumption, we use the midpoint $N_k + \dfrac{\eta}{2}$ to represent the $k$th bin $[N_k, N_k + \eta)$.
\\\\
We approximate the theoretical mean $\E\Big[\Delta N \Big| N = N_k + \dfrac{\eta}{2}\Big]$ with the empirical mean $\Big\langle\Delta N \Big| N = N_k + \eta_i, 0 \leq \eta_i < \eta, \hat{S}_k\Big\rangle$ and the theoretical variance $\V\Big[\Delta N \Big| N = N_k + \dfrac{\eta}{2}\Big]$ the empirical variance $\sigma^2\Big[\Delta N \Big| N = N_k + \eta_i, 0 \leq \eta_i < \eta, \hat{S}_k\Big]$ obtained from simulation of $S$ cell number trajectories. 
Recall that $\hat{S}_k$ denotes the number of population size $N_k + \eta_i$ landing in bin $k$.
These sample sizes $\hat{S}_k, k = 1,2,\ldots,k_{\max}$, are not necessarily equal to each other or equal to the number of cell number trajectories $S$, which is pre-determined and independent of the bin size $\eta$.
Different bin sizes $\eta$ result in different sets of $\hat{S}_k, k = 1,2,\ldots,k_{\max}$. With the same bin size $\eta$, different simulations may also result in different sets of cell number values and hence different sets of $\hat{S}_k, k = 1,2,\ldots,k_{\max}$.
It is well-known that as the larger the sample size $\hat{S}_k$, the smaller the estimation errors \parencite{Ferlic:2019}.
\\\\
In this section, we analyze how the bin size influences distributions of estimation errors of birth and death rates. In particular, we compute the theoretical means and variances of errors as functions of bin size $\eta$. We use the notation $N$ for cell number to be consistent with the mathematical model discussed in Section \ref{sect:model}. A summary of notations can be found in Section \ref{appendix:notation}.

\subsection{Theoretical Mean and Variance of Cell Number Increment as Functions of Bin Size}
\begin{figure}[H]
  \centering
  \includegraphics[scale=0.75]{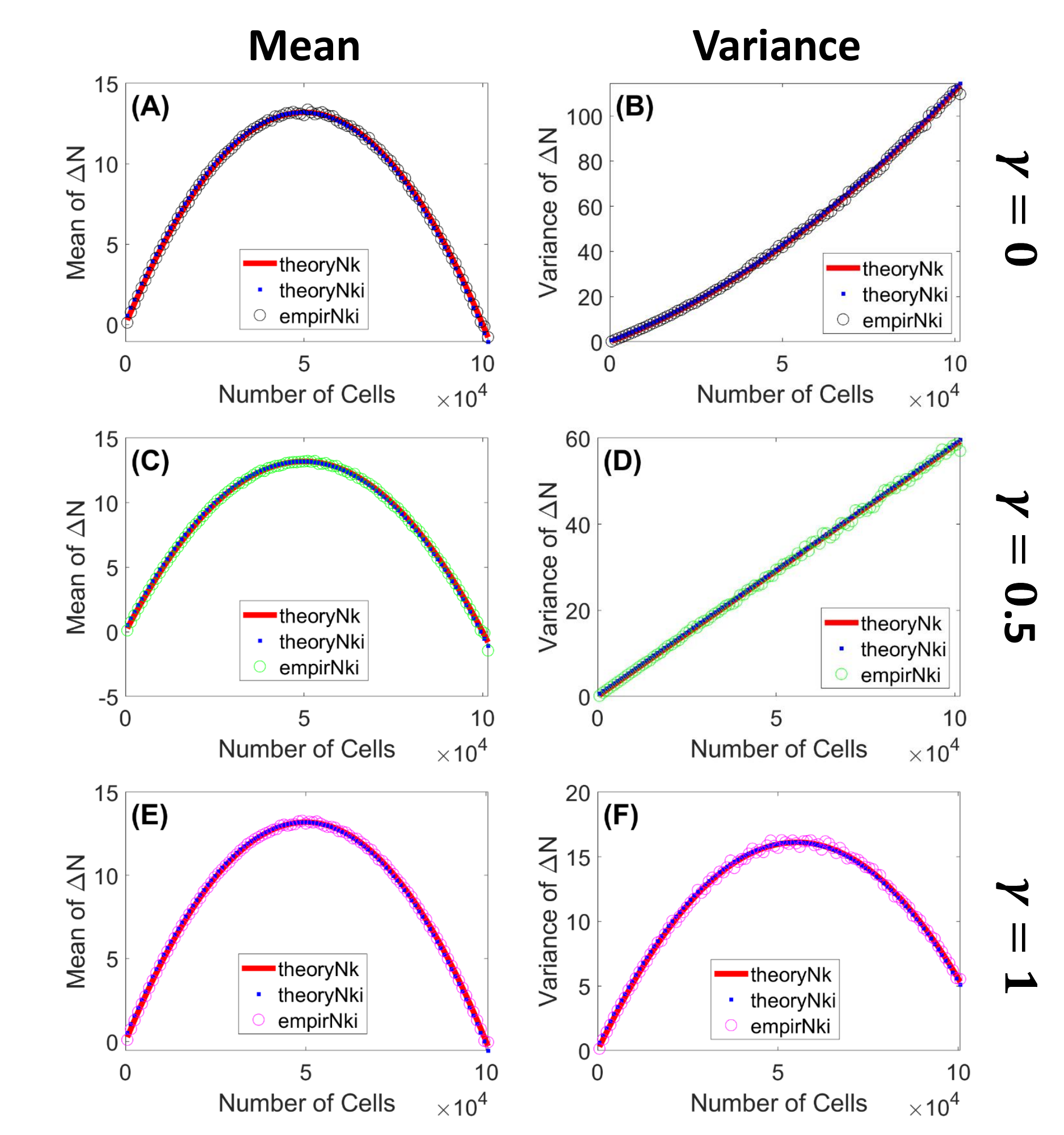}
  \caption{\textbf{Theoretical mean and variance of cell number increments $\Delta N$ as functions of population size are well-aligned with empirical mean and variance}. The statistics are computed using carrying capacity $K = 10^5$ and bin size $\eta = 10^3$. In \textbf{(A, C, E)}, we compare the theoretical mean $\E\Big[\Delta N \Big| N = N_k + U, U\sim \text{Unif}[0,\eta)\Big]$ with the theoretical mean $\E\Big[\Delta N \Big| N = N_k + \dfrac{\eta}{2}\Big]$ and the empirical mean $\Big\langle\Delta N \Big| N = N_k + \eta_i, 0 \leq \eta_i < \eta\Big\rangle$ using data from a simulation of $S = 100$ cell number trajectories. In \textbf{(B, E, F)}, we compare the theoretical variance $\V\Big[\Delta N \Big| N = N_k + U, U\sim \text{Unif}[0,\eta)\Big]$ with theoretical variance $\V\Big[\Delta N \Big| N = N_k + \dfrac{\eta}{2}\Big]$ and the empirical variance $\sigma^2\Big[\Delta N \Big| N = N_k + \eta_i, 0 \leq \eta_i < \eta\Big]$ using data from a simulation of $S = 100$ cell number trajectories. \textbf{(A, B)}: $\gamma = 0$ (black color); \textbf{(C, D)}: $\gamma = 0.5$ (green color); \textbf{(E, F)}: $\gamma = 0.5$ (magenta color). Red lines (-) denote theoretical statistics (i.e.~mean and variance) for $N = N_k + \dfrac{\eta}{2}$; blue squares ($\square$) denote theoretical statistics for $N = N_k + U, U\sim \text{Unif}[0,\eta)$; circles (◦) denote empirical statistics for $N = N_k + \eta_i$ with $i = 1, \ldots, \hat{S}_k$. } 
  \label{fig:erroranalysis-meanvar-DeltaN}
\end{figure}

As mentioned above, our estimation of the birth and death rates corresponding to $N=N_k+\dfrac{\eta}{2}$ uses the empirical mean $\Big\langle\Delta N \Big| N = N_k + \eta_i, 0 \leq \eta_i < \eta, \hat{S}_k\Big\rangle$ and empirical variance $\sigma^2\Big[\Delta N|N = N_k + \eta_i]\Big|0 \leq \eta_i < \eta, \hat{S}_k \Big]$.
The theoretical means and variances of the estimation errors involves the theoretical mean $\E\Big[\Delta N \Big| N = N_k + U, U\sim \text{Unif}[0,\eta)\Big]$ and theoretical variance $\V\Big[\Delta N \Big| N = N_k + U, U\sim \text{Unif}[0,\eta)\Big]$, as shown in Section \ref{sect:valid-error-analysis}.
In this subsection, we analyze how the bin size $\eta$ influences these theoretical mean and variance. We present the analysis for nonnegative birth rates, that is, in which we can drop the $\max$ function in Equation \eqref{eq:birth-rate-definition}, as the birth rates are always positive in our simulated datasets.
\\\\
Theoretical mean:
\begin{align}
    &\E\Big[\Delta N \Big| N = N_k + U, U\sim \text{Unif}[0,\eta)\Big] = \E\Big[\E[\Delta N|N = N_k + U]\Big|U\sim \text{Unif}[0,\eta) \Big]\\
    =& \E\Big[(b_{N_k + U} - d_{N_k + U})(N_k + U)\Delta t\Big|U\sim \text{Unif}[0,\eta)\Big]\\
    =& \E\Big[(r- \dfrac{r}{K}N_k - \dfrac{r}{K}U)(N_k + U)\Delta t\Big| U\sim \text{Unif}[0,\eta)\Big]\\
    =& \E\Big[(r - \dfrac{r}{K}N_k)N_k\Delta t\Big| U\sim \text{Unif}[0,\eta)\Big] - \dfrac{r}{K}N_k\Delta t\E\Big[U\Big|U\sim \text{Unif}[0,\eta)\Big]\\
    &+ (r - \dfrac{r}{K}N_k)\Delta t\E\Big[U \Big| U\sim \text{Unif}[0,\eta)\Big] - \dfrac{r}{K}\Delta t \E\Big[U^2\Big|U\sim \text{Unif}[0,\eta) \Big]\nonumber\\
    =& \E[\Delta N\given N=N_k] + (r - 2\dfrac{r}{K}N_k)\Delta t\dfrac{\eta}{2} - \dfrac{r}{K}\Delta t \dfrac{\eta^2}{3}.
\end{align}
Theoretical variance:
\begin{align}
    &\V\Big[\Delta N \Big| N = N_k + U, U\sim \text{Unif}[0,\eta)\Big]\nonumber\\
    =& \E\Big[\Delta N^2 \Big| N = N_k + U, U\sim \text{Unif}[0,\eta)\Big] - \Big(\E\Big[\Delta N \Big| N = N_k + U, U\sim \text{Unif}[0,\eta)\Big]\Big)^2,
\end{align}
where
\begin{align}
    &\E\Big[\E[\Delta N^2|N = N_k + U]\Big|U\sim \text{Unif}[0,\eta) \Big]\nonumber\\
    =& \E\Bigg[\V[\Delta N|N = N_k + U] + \Big(\E[\Delta N|N = N_k + U]\Big)^2\Big|U\sim \text{Unif}[0,\eta) \Bigg]\\
    =& \E\Big[\V[\Delta N|N = N_k + U]\Big|U\sim \text{Unif}[0,\eta)\Big] + \E\Big[\Big(\E[\Delta N|N = N_k + U]\Big)^2\Big|U\sim \text{Unif}[0,\eta)\Big],
\end{align}
and
\begin{align}
    &\E\Big[\V[\Delta N|N = N_k + U]\Big|U\sim \text{Unif}[0,\eta]\Big] \nonumber\\
    &= \E\Big[(b_{N_k + U} + d_{N_k + U})(N_k + U)\Delta t\Big| U\sim \text{Unif}[0,\eta)\Big]\\
    &= \E\Big[(b_0 + d_0 + (1-2\gamma)\dfrac{r}{K}N_k + (1-2\gamma)\dfrac{r}{K}U)(N_k + U)\Delta t\Big| U\sim \text{Unif}[0,\eta)\Big]\\
    &= \E\Big[(b_0 + d_0 + (1-2\gamma)\dfrac{r}{K}N_k)N_k\Delta t\Big| U\sim \text{Unif}[0,\eta)\Big] +  \E\Big[(b_0 + d_0 + (1-2\gamma)\dfrac{r}{K}N_k)U\Delta t\Big| U\sim \text{Unif}[0,\eta)\Big]\\
    &+ \E\Big[(1-2\gamma)\dfrac{r}{K}N_kU\Delta t\Big| U\sim \text{Unif}[0,\eta)\Big] + \E\Big[(1-2\gamma)\dfrac{r}{K}U^2\Delta t\Big| U\sim \text{Unif}[0,\eta)\Big]\nonumber\\
    &= \V[\Delta N_k] + \Big(b_0 + d_0 + 2(1-2\gamma)\dfrac{r}{K}N_k\Big)\Delta t\E[U | U\sim \text{Unif}[0,\eta)] + (1-2\gamma)\dfrac{r}{K}\Delta t \E[U^2 | U\sim \text{Unif}[0,\eta)]\\
    &= \V[\Delta N_k] + \Big(b_0 + d_0 + 2(1-2\gamma)\dfrac{r}{K}N_k\Big)\Delta t\dfrac{\eta}{2} + (1-2\gamma)\dfrac{r}{K}\Delta t \dfrac{\eta^2}{3},
\end{align}
and
\begin{align}
    &\E\Big[(b_{N_k + U} - d_{N_k + U})^2(N_k + U)^2\Delta t^2\Big|U\sim \text{Unif}[0,\eta) \Big]\nonumber\\
    &=\Delta t^2  \E\Big[\Big((b_0 - d_0)(N_k + U) - \dfrac{r}{K}(N_k + U)^2\Big)^2\Big| U\sim \text{Unif}[0,\eta)\Big]\\
    &=\Delta t^2  \E\Big[r^2(N_k + U)^2 - 2\dfrac{r^2}{K}(N_k + U)^3 + \dfrac{r^2}{K^2}(N_k + U)^4\Big| U\sim \text{Unif}[0,\eta)\Big]\\
    &= \Delta t^2 r^2 \E\Big[(N_k + U)^2\Big| U\sim \text{Unif}[0,\eta) \Big] - 2\dfrac{r^2}{K}\Delta t^2\E\Big[(N_k + U)^3\Big| U\sim \text{Unif}[0,\eta) \Big]\\
    &+ \dfrac{r^2}{K^2}\Delta t^2 \E\Big[(N_k + U)^4\Big| U\sim \text{Unif}[0,\eta) \Big]\nonumber\\
    &= \Delta t^2 r^2 \dfrac{(N_k + \eta)^3 - N_k^3}{3\eta} - 2\dfrac{r^2}{K}\Delta t^2 \dfrac{(N_k + \eta)^4 - N_k^4}{4\eta} + \dfrac{r^2}{K^2}\Delta t^2\dfrac{(N_k + \eta)^5 - N_k^5}{5\eta}.
\end{align}
Therefore,
\begin{align}
    &\V\Big[\Delta N \Big| N = N_k + U, U\sim \text{Unif}[0,\eta)\Big] \nonumber\\
    =& \V[\Delta N_k] + \Big(b_0 + d_0 + 2(1-2\gamma)\dfrac{r}{K}N_k\Big)\Delta t\dfrac{\eta}{2} + (1-2\gamma)\dfrac{r}{K}\Delta t \dfrac{\eta^2}{3}\label{eq:true_vardNki}\\
    &+ \Delta t^2 r^2 \dfrac{(N_k + \eta)^3 - N_k^3}{3\eta} - 2\dfrac{r^2}{K}\Delta t^2 \dfrac{(N_k + \eta)^4 - N_k^4}{4\eta} + \dfrac{r^2}{K^2}\Delta t^2\dfrac{(N_k + \eta)^5 - N_k^5}{5\eta}\nonumber\\ 
    &- \Bigg(\E[\Delta N\given N=N_k] + (r - 2\dfrac{r}{K}N_k)\Delta t\dfrac{\eta}{2} - \dfrac{r}{K}\Delta t \dfrac{\eta^2}{3} \Bigg)^2\nonumber.
\end{align}
In Figure \ref{fig:erroranalysis-meanvar-DeltaN}, we compare the theoretical mean $\E\Big[\Delta N \Big| N = N_k + U, U\sim \text{Unif}[0,\eta)\Big]$ that we just computed with the theoretical mean $\E\Big[\Delta N \Big| N = N_k + \dfrac{\eta}{2}\Big]$ and the empirical mean $\Big\langle\Delta N \Big| N = N_k + U, U\sim \text{Unif}[0,\eta)\Big\rangle$ using data from a simulation of $S = 100$ cell number trajectories. Similarly, we also compare the population variance $\V\Big[\Delta N \Big| N = N_k + U, U\sim \text{Unif}[0,\eta)\Big]$ that we just computed with the theoretical variance $\V\Big[\Delta N \Big| N = N_k + \dfrac{\eta}{2}\Big]$ and the empirical variance $\sigma^2\Big[\Delta N \Big| N = N_k + \eta_i, 0 \leq \eta_i < \eta\Big]$ using data from a simulation of $S = 100$ cell number trajectories.
\\\\
\subsection{Errors of Birth and Death Rate Estimation as Functions of Bin Size}
In this section we consider the effect of bin size on the accuracy with which we can estimate the birth and death rates.
Thus we compare the theoretical mean and variance of the population increment, given that a point of the trajectory lies within a given bin, versus the empirical mean and variance obtained from simulation with a finite sample size.  
We use $\mathcal{E}$ to represent expected differences in these errors.
Define
\begin{align}
    \mathcal{E}_{k\text{mean}} &:= \E\Big[\Delta N \Big| N = N_k + \dfrac{\eta}{2}\Big] - \Big\langle\Delta N \Big| N = N_k + \eta_i, 0 \leq \eta_i < \eta, \hat{S}_k\Big\rangle,\\
     \mathcal{E}_{k\text{var}} &:= \V\Big[\Delta N \Big| N = N_k + \dfrac{\eta}{2}\Big] - \sigma^2\Big[\Delta N \Big| N = N_k + \eta_i, 0 \leq \eta_i < \eta,\hat{S}_k\Big].
\end{align}
The errors in estimating the birth and death rates corresponding to $N = N_k + \dfrac{\eta}{2}$ are
\begin{align}
    \mathcal{E}_{k\text{birth}} = \dfrac{\mathcal{E}_{k\text{var}} + \mathcal{E}_{k\text{mean}} }{2\Delta t} \quad \text{ and } \quad \mathcal{E}_{k\text{death}} = \dfrac{\mathcal{E}_{k\text{var}} - \mathcal{E}_{k\text{mean}} }{2\Delta t}. \label{eq:errors-expected}
\end{align}
The theoretical means of the errors over \textit{all} realizations $\eta_i$ of the iid uniform random variable $U\sim\text{Unif}[0,\eta)$ are
\begin{align}
    \E\Big[\mathcal{E}_{k\text{birth}}\Big] = \dfrac{\E\Big[\mathcal{E}_{k\text{var}}\Big] + \E\Big[\mathcal{E}_{k\text{mean}}\Big] }{2\Delta t} \quad \text{ and } \quad \E\Big[\mathcal{E}_{k\text{death}}\Big] = \dfrac{\E\Big[\mathcal{E}_{k\text{var}}\Big] - \E\Big[\mathcal{E}_{k\text{mean}}\Big] }{2\Delta t}. \label{eq:errors-variance}
\end{align}
The theoretical variances of the errors over \textit{all} realizations $\eta_i$ of $U$ are
\begin{align}
    \V\Big[\mathcal{E}_{k\text{birth}}\Big] = 
    \V\Big[\mathcal{E}_{k\text{death}}\Big] = 
    \dfrac{\V\Big[\mathcal{E}_{k\text{var}}\Big] + \V\Big[\mathcal{E}_{k\text{mean}}\Big] }{4\Delta t^2}.
\end{align}
We analyze how the bin size $\eta$ influences these analytical expected values and variances of errors $\E\Big[\mathcal{E}_{k\text{mean}}\Big]$, $\E\Big[\mathcal{E}_{k\text{var}}\Big]$,
$\V\Big[\mathcal{E}_{k\text{var}}\Big]$, and $\V\Big[\mathcal{E}_{k\text{mean}}\Big]$.\\\\
Treating the samples of $\Big(\Delta N \Big| N = N_k + U, U \sim \text{Unif}[),\eta)\Big)$ as if they were identically and independently distributed, the expected value of the sample mean is equal to the theoretical mean. Therefore,
\begin{align}
    \E\Big[\mathcal{E}_{k\text{mean}}\Big] &= \E\Bigg[\underbrace{\E\Big[\Delta N \Big| N = N_k + \dfrac{\eta}{2}\Big]}_{\text{independent of $\eta_i$}}\Bigg] - \E\Bigg[\Big\langle\Delta N \Big| N = N_k + \eta_i, 0 \leq \eta_i < \eta, \hat{S}_k\Big\rangle\Bigg]\\
    &= \E\Big[\Delta N \Big| N = N_k + \dfrac{\eta}{2}\Big] - \E\Big[\Delta N \Big| N = N_k + U, U\sim \text{Unif}[0,\eta)\Big],
\end{align}
where
\begin{align}
    &\E\Big[\Delta N \Big| N = N_k + \dfrac{\eta}{2}\Big]\nonumber\\
    &= \Big(b_{N_k + (\eta/2)} - d_{N_k + (\eta/2)}\Big)\Big(N_k + \dfrac{\eta}{2}\Big)\Delta t\\
    &= \Big(r-\dfrac{r}{K}N_k -\dfrac{r}{K}\dfrac{\eta}{2}\Big)\Big(N_k + \dfrac{\eta}{2}\Big)\Delta t\\
    &= \Big(r-\dfrac{r}{K}N_k\Big)N_k\Delta t + \Big(r-\dfrac{r}{K}N_k\Big)\Delta t\dfrac{\eta}{2} - \dfrac{r}{K}N_k\Delta t\dfrac{\eta}{2} - \dfrac{r}{K}\dfrac{\eta^2}{4}\Delta t\\
    &= \E[\Delta N| N = N_k] + \Big( r-2\dfrac{r}{K}N_k\Big)\Delta t\dfrac{\eta}{2} - \dfrac{r}{K}\Delta t\dfrac{\eta^2}{4}.
\end{align}
Hence,
\begin{align}
    \E\Big[\mathcal{E}_{k\text{mean}}\Big] =& \E[\Delta N| N = N_k] + \Big( r-2\dfrac{r}{K}N_k\Big)\Delta t\dfrac{\eta}{2} - \dfrac{r}{K}\Delta t\dfrac{\eta^2}{4}\\
    &- \E[\Delta N\given N=N_k] - (r - 2\dfrac{r}{K}N_k)\Delta t\dfrac{\eta}{2} + \dfrac{r}{K}\Delta t \dfrac{\eta^2}{3}\nonumber\\
    =& \dfrac{1}{12}\dfrac{r}{K}\Delta t \eta^2. \label{eq:error-mean-expected}
\end{align}
We observe that the expected error $\E\Big[\mathcal{E}_{k\text{mean}}\Big]$ in approximating the true mean $\E\Big[\Delta N \Big| N = N_k + \dfrac{\eta}{2}\Big]$ for each bin $k$ is independent of $k$ and is increasing quadratically for $\eta > 0$. 
If we write the expected error $\E\Big[\mathcal{E}_{k\text{mean}}\Big]$ as $\dfrac{1}{12}\Big(r\Delta t\Big)\Big(\dfrac{\eta}{K}\Big)\eta$, then we see that the expected error depends on the ratio $\Big(\dfrac{\eta}{K}\Big)$, which shows how big the bin size is relative to the system size (i.e. carrying capacity $K$), and also depends on the product $r\Delta t$, which can be interpreted roughly as the \textit{per capita} change in cell number $\Big(\dfrac{\Delta N}{N}\Big)$ after $\Delta t$. 
The higher these ratios are, the higher expected error is.
Looking from a different angle, the expected error $\E\Big[\mathcal{E}_{k\text{mean}}\Big]$ can be written as $\Big(\dfrac{r\Delta t}{K}\Big)\Big(\dfrac{1}{12}\eta^2\Big)$. 
This shows that the expected error depends on $\Big(\dfrac{1}{12}\eta^2 \Big)$, which is the variance of the random variable $\eta_i$, and how big the  \textit{per capita} change in cell number $r\Delta t$ after $\Delta t$ is relative to the system size $K$. This observation suggests that it may be harder to estimate the cell number increments with high accuracy for fast-reproducing cell types. Further analysis on the relation between $r$ and $K$ would be interesting for future work, since existing work such as \parencite{reding2017unconstrained} shows that the product $rK$ can influence the evolution of antibiotic-resistant bacterial genomes.
\\\\
We assume that the samples of $\Big(\Delta N \Big| N = N_k + U, U\sim \text{Unif}[0,\eta)\Big)$ are independently and identically distributed, so the expected value of the sample variance is equal to the population variance. Therefore,
\begin{align}
    \E\Big[\mathcal{E}_{k\text{var}}\Big] =& \E\Bigg[\underbrace{\V\Big[\Delta N \Big| N = N_k + \dfrac{\eta}{2}\Big]}_{\text{independent of $\eta_i$}}\Bigg] - \E\Bigg[\sigma^2\Big[\Delta N \Big| N = N_k + \eta_i, 0 \leq \eta_i < \eta,\hat{S}_k\Big] \Bigg]\\
    =& \V\Big[\Delta N \Big| N = N_k + \dfrac{\eta}{2}\Big] - \V\Big[\Delta N \Big| N = N_k + U, U\sim \text{Unif}[0,\eta)\Big],
\end{align}
where
\begin{align}
   &\V\Big[\Delta N \Big| N = N_k + \dfrac{\eta}{2}\Big]\nonumber\\
    =& \Big(b_0 + d_0 + (1-2\gamma)\dfrac{r}{K}N_k + (1-2\gamma)\dfrac{r}{K}\dfrac{\eta}{2} \Big)\Big(N_k + \dfrac{\eta}{2}\Big)\Delta t\\
    =&\V[\Delta N_k] + \Big(b_0 + d_0 + 2(1-2\gamma)\dfrac{r}{K}N_k\Big)\Delta t\dfrac{\eta}{2} + (1-2\gamma)\dfrac{r}{K}\Delta t \dfrac{\eta^2}{4}.
\end{align}
Therefore,
\begin{align}
    \E\Big[\mathcal{E}_{k\text{var}}\Big] =& -(1-2\gamma)\dfrac{r}{K}\Delta t \dfrac{\eta^2}{12}\\
    &+ \Delta t^2 r^2 \dfrac{(N_k + \eta)^3 - N_k^3}{3\eta} - 2\dfrac{r^2}{K}\Delta t^2 \dfrac{(N_k + \eta)^4 - N_k^4}{4\eta} + \dfrac{r^2}{K^2}\Delta t^2\dfrac{(N_k + \eta)^5 - N_k^5}{5\eta}\nonumber\\ 
    &- \Bigg(\E[\Delta N\given N=N_k] + (r - 2\dfrac{r}{K}N_k)\Delta t\dfrac{\eta}{2} - \dfrac{r}{K}\Delta t \dfrac{\eta^2}{3} \Bigg)^2.\nonumber
\end{align}
Now, we compute the theoretical variances $\V\Big[\mathcal{E}_{k\text{mean}}\Big]$ and $\V\Big[\mathcal{E}_{k\text{var}}\Big]$ over \textit{all} realizations of $\eta_i$. We assume the samples of $\Big(\Delta N \Big| N = N_k + U, U \sim \text{Unif}[0,\eta) \Big)$ are identically distributed, so the variance of the sample mean is equal to the population variance divided by the sample size. Therefore,
\begin{align}
\V\Big[\mathcal{E}_{k\text{mean}}\Big] &= \V\Bigg[\underbrace{\E\Big[\Delta N \Big| N = N_k + \dfrac{\eta}{2}\Big]}_{\text{independent of $\eta_i$}}\Bigg] + \V\Bigg[\Big\langle\Delta N \Big| N = N_k + U, 0 \leq \eta_i < \eta, \hat{S}_k\Big\rangle\Bigg]\\ 
&= \V\Bigg[\Big\langle\Delta N \Big| N = N_k + \eta_i, 0 \leq \eta_i < \eta, \hat{S}_k\Big\rangle\Bigg]\\
&= \dfrac{\V\Big[\Delta N \Big| N = N_k + U, U\sim \text{Unif}[0,\eta) \Big]}{\hat{S}_k}.
\end{align}
As mentioned above, the samples of $\Big(\Delta N \Big| N = N_k + U, U \sim \text{Unif}[0,\eta)\Big)$ are independently and identically distributed. For computation convenience here, we approximate the binomial distribution of these samples with the Gaussian distribution with the empirical mean and variance as discussed in Section \ref{sect:data-simulation}. We still use the notation $N$ instead of $X$ here to be consistent with the other statistics computed above. With this approximation, the theoretical variance of the empirical variance is equal to two times the theoretical variance squared divided by the sample size minus one. Therefore,
\begin{align}
    \V\Big[\mathcal{E}_{k\text{var}}\Big] =& \V\Bigg[\underbrace{\V\Big[\Delta N \Big| N = N_k + \dfrac{\eta}{2}\Big]}_{\text{independent of $\eta_i$}}\Bigg] + \V\Bigg[\sigma^2\Big[\Delta N \Big| N = N_k + \eta_i, 0 \leq \eta_i < \eta,\hat{S}_k\Big]\Bigg]\\
    =&\V\Bigg[\sigma^2\Big[\Delta N \Big| N = N_k + \eta_i, 0 \leq \eta_i < \eta,\hat{S}_k\Big]\Bigg]\\
    =&\dfrac{2\Big(\V\Big[\Delta N \Big| N = N_k + U, U\sim \text{Unif}[0,\eta)\Big]\Big)^2}{\hat{S}_k-1}.
\end{align}
The theoretical variance $\V\Big[\Delta N \Big| N = N_k + U, U\sim \text{Unif}[0,\eta)\Big]$ is given by Equation \eqref{eq:true_vardNki}. 
\\\\
Using the $\E\Big[\mathcal{E}_{k\text{mean}}\Big]$, $\E\Big[\mathcal{E}_{k\text{var}}\Big]$, $\V\Big[\mathcal{E}_{k\text{mean}}\Big]$, and $\V\Big[\mathcal{E}_{k\text{var}}\Big]$ that we just computed, we obtain the theoretical means and variances of the errors in estimating birth and death rates corresponding to $N=N_k + \dfrac{\eta}{2}$ for all $k = 1,2,\ldots,k_{\max}$ using Equations \eqref{eq:errors-expected} and \eqref{eq:errors-variance}.
\\\\
In Figure \ref{fig:erroranalysis-errorbirthdeath}, we compare the $l_2$-norm of the theoretical means and variances of the errors and compare them with the $l_2$-norm of the empirical errors (i.e. realizations of the error random variables) computed using data from a simulation of $S=100$ cell number trajectories. To computed the theoretical variances of the errors shown in Figure \ref{fig:erroranalysis-errorbirthdeath}, we use the empirical sample sizes $\hat{S}_k, k = 1,2,\ldots,k_{\max}$, from the same data simulation.  
\\\\
\noindent
We observe that as the bin size $\eta$ increases, the theoretical means of the errors increase, the theoretical variances (or standard deviations) of the errors decreases, and the empirical errors balance between the theoretical means and variances (or standard deviations) and have convex quadratic shapes. 
The theoretical means of the errors reflect the differences between $\Delta N$ at one point $\Big(N = N_k + \dfrac{\eta}{2}\Big)$ and $\Delta N$ at multiple points $\Big(N = N_k + \eta_i, 0 \leq \eta_i < \eta\Big)$; the smaller the bin size, the closer multiple points are to one point, so the error is smaller (for example, Equation \eqref{eq:error-mean-expected} shows that the expected errors in estimating the mean of cell number increments are $(r\Delta t/12K)\eta^2$). 
However, if the bin is too small, then there are not enough samples to estimate theoretical statistics with empirical statistics with accuracy. 
The theoretical variances of errors involves sample sizes; the bigger the bin size, the more samples we have. 
These two competing effects of bin size result in the empirical errors being intermediate values between the two theoretical statistics (means and variances) of the estimation errors. 
The optimal bin size reflects a balancing of these two effects.
When the bin size is smaller than the optimal bin size, the sample error coincides with the sum of the expected error and the standard deviation of the error.
When the bin size is bigger than the optimal bin size, this relationship breaks down, which may reflect growing inaccuracy of our approximation that the trajectory points are uniformly and i.i.d.~within each bin. 
\subsection{Notation}\label{appendix:notation}
\begin{align*}
    N &\quad \text{ denotes } \quad \text{discrete cell number random variable}\\
    t_0 \text{ and } t_T &\quad \text{ denotes } \quad \text{deterministic initial and final times respectively}\\
    \eta &\quad \text{ denotes } \quad \text{deterministic bin size}\\
    k &\quad \text{ denotes } \quad \text{bin index, $k = 1, 2, \ldots, k_{\max}$}\\
    U &\quad \text{ denotes } \quad \text{uniformly distributed random variable such that $\Big(N=N_k+U\Big) \in [N_k, N_k+\eta)$}\\
    \eta_i &\quad \text{ denotes } \quad \text{realization of the random variable $U$}\\
    S &\quad \text{ denotes } \quad \text{number of cell number trajectories/time series}\\
    \hat{S}_k &\quad \text{ denotes } \quad \text{number of samples of $\Delta N:= N(t+\Delta t) - N(t)$ in bin $[N_k, N_k + \eta)$}\\
    \E[\cdot] &\quad \text{ denotes } \quad \text{theoretical mean}\\
    \langle \cdot \rangle &\quad \text{ denotes } \quad \text{empirical mean}\\
    \V[\cdot] &\quad \text{ denotes } \quad \text{theoretical variance}\\
    \sigma^2[\cdot] &\quad \text{ denotes } \quad \text{empirical variance}\\
    \mathcal{E}_{[\cdot]} &\quad \text{ denotes } \quad \text{error}\\
\end{align*}

\section{Analysis of Log-Likelihood Function}\label{appendix:appx-log-likelihood}
We calculate the first and second derivatives of the log-likelihood function $f(\gamma)$  \eqref{eq:log-likelihood} for a single trajectory as a function of the density dependence parameter $\gamma$. 
Let $\Delta x_j = x_{j+1}-x_j$.
\begin{align}
    f(\gamma) &= \sum_{j=1}^{T-1}\dfrac{1}{2}\ln\left(\dfrac{1}{2\pi \V[\Delta x_j]}\right) - \dfrac{1}{2}\dfrac{\Big(\Delta x_j - \E[\Delta x_j]\Big)^2}{\V[\Delta x_j]}\\
    &= \sum_{j=1}^{T-1} -\dfrac{1}{2}\ln(2\pi) - \dfrac{1}{2}\ln \Big(\V[\Delta x_j] \Big) - \dfrac{1}{2}\dfrac{\Big(\Delta x_j - \E[\Delta x_j]\Big)^2}{\V[\Delta x_j]},
\end{align}
where
\begin{align}
    \E[\Delta x_j] = x_j\Delta t (b_{x_j} - d_{x_j}) = x_j\Delta t \Bigg(\dfrac{b_0 - \gamma(r/K)x_j + \Big|b_0 - \gamma(r/K)x_j  \Big|}{2} - d_0 - (1-\gamma)(r/K)x_j \Bigg),\\
    \V[\Delta x_j] = x_j\Delta t (b_{x_j} + d_{x_j}) = 2x_j\Delta t \Bigg(\dfrac{b_0 - \gamma(r/K)x_j + \Big|b_0 - \gamma(r/K)x_j  \Big|}{2} + d_0 + (1-\gamma)(r/K)x_j \Bigg).
\end{align}
We observe that $\E[\Delta x_j]$ is a piecewise linear function of $\gamma$, i.e. $\E[\Delta x_j]$ has the form $c_1^j + c_2^j\gamma$, where
\begin{align}   
    c_1^j = 
    \begin{cases}
    x_j\Delta t\Big(b_0 - d_0 - \dfrac{r}{K}x_j\Big), &\text{ for } \Big(b_0 - \gamma\dfrac{r}{K}x_j\Big) > 0,\\
    -x_j\Delta t\Big(d_0 + \dfrac{r}{K}x_j\Big), &\text{ for } \Big(b_0 - \gamma\dfrac{r}{K}x_j^2\Big) = 0,
    \end{cases}
\end{align}
and
\begin{align}   
    c_2^j = 
    \begin{cases}
    0, &\text{ for } \Big(b_0 - \gamma\dfrac{r}{K}x_j\Big) > 0,\\ 
    \Delta t\dfrac{r}{K}x_j, &\text{ for } \Big(b_0 - \gamma\dfrac{r}{K}x_j^2\Big) = 0.
    \end{cases}
\end{align}
The variance $\V[\Delta x_j]$ is also a linear function of $\gamma$, i.e. $\V[\Delta x_j]$ has the form $c_3^j - c_4^j\gamma$ with
\begin{align}   
    c_3^j = 
    \begin{cases}
    x_j\Delta t\Big(b_0 + d_0 + \dfrac{r}{K}x_j \Big), &\text{ for } \Big(b_0 - \gamma\dfrac{r}{K}x_j\Big) > 0,\\ 
    x_j\Delta t\Big(d_0 + \dfrac{r}{K}x_j \Big), &\text{ for } \Big(b_0 - \gamma\dfrac{r}{K}x_j^2\Big) = 0,
    \end{cases}
\end{align}
and 
\begin{align}   
    c_4^j = 
    \begin{cases}
    2\Delta t \dfrac{r}{K}x_j^2, &\text{ for } \Big(b_0 - \gamma\dfrac{r}{K}x_j\Big) > 0,\\ 
    \Delta t \dfrac{r}{K}x_j^2, &\text{ for } \Big(b_0 - \gamma\dfrac{r}{K}x_j^2\Big) = 0,
    \end{cases}
\end{align}
Therefore,
\begin{align}
    f(\gamma) =\sum_{j=1}^{T-1} -\dfrac{1}{2}\ln(2\pi) - \dfrac{1}{2}\ln \Big(c_3^j - c_4^j\gamma \Big) - \dfrac{1}{2}\dfrac{\Big(\Delta x_j - c_1^j-c_2^j\gamma\Big)^2}{c_3^j-c_4^j\gamma}.
\end{align}
Denote $v_j = \dfrac{1}{c_3^j - c_4^j\gamma} \Rightarrow v_j > 0$ and $\dfrac{dv_j}{d\gamma} = \dfrac{c_4^j}{(c_3^j - c_4^j\gamma)^2} = c_4^jv_j^2$. We have
\begin{align}
    f(\gamma) = \sum_{j=1}^{T-1} -\dfrac{1}{2}\ln(2\pi) + \dfrac{1}{2}\ln(v_j) - \dfrac{1}{2}\Big(\Delta x_j - c_1^j-c_2^j\gamma\Big)^2 v_j.
\end{align}
If $b_0 - \gamma(r/K)x_j > 0$, then $c_2^j = 0$ and $\E[\Delta x_j] = c_1^j$ is independent of $\gamma$. Hence,
\begin{align}
    \dfrac{df}{d\gamma} =& \sum_{j=1}^{T-1} \dfrac{1}{2}\dfrac{1}{v_j}\dfrac{dv_j}{d\gamma} - \dfrac{1}{2}\Big(\Delta x_j - c_1^j\Big)^2\dfrac{dv_j}{d\gamma} = \sum_{j=1}^{T-1} \dfrac{1}{2}c_4^j v_j - \dfrac{1}{2}\Big(\Delta x_j - c_1^j\Big)^2c_4^jv_j^2\\
    \Rightarrow \dfrac{d^2f}{d\gamma^2} =& \sum_{j=1}^{T-1} \dfrac{1}{2}c_4^j\dfrac{dv_j}{d\gamma} - \dfrac{1}{2}\Big(\Delta x_j - c_1^j\Big)^2c_4^j2v_j\dfrac{dv_j}{d\gamma}\\
    =& \sum_{j=1}^{T-1}\dfrac{1}{2}(c_4^j)^2v_j^2 - \Big(\Delta x_j - c_1^j\Big)^2(c_4^j)^2v_j^3\\
    =& \sum_{j=1}^{T-1}(c_4^j)^2v_j^2\Bigg(\dfrac{1}{2} - \Big(\Delta x_j - c_1^j)\Big)^2v_j  \Bigg)\\
    =& \sum_{j=1}^{T-1}(c_4^j)^2v_j^2\Bigg(\dfrac{1}{2} - \dfrac{(\Delta x_j - \E[\Delta x_j])\Big)^2}{\V[\Delta x_j]}  \Bigg).
\end{align}
In general,
\begin{align}
    \dfrac{df}{d\gamma} &= \sum_{j=1}^{T-1} \dfrac{1}{2}\dfrac{1}{v_j}\dfrac{dv_j}{d\gamma} - \dfrac{1}{2}\Big(\Delta x_j - c_1^j - c_2^j\gamma\Big)^2\dfrac{dv_j}{d\gamma} + c_2^j\Big(\Delta x_j - c_1^j - c_2^j\gamma\Big)v_j\\
    =& \sum_{j=1}^{T-1} \dfrac{1}{2}c_4^jv_j - \dfrac{1}{2}\Big(\Delta x_j - c_1^j - c_2^j\gamma\Big)^2c_4^jv_j^2 + c_2^j \Big(\Delta x_j - c_1^j - c_2^j\gamma\Big)v_j\\
\Rightarrow \dfrac{d^2f}{d\gamma^2} 
    =& \sum_{j=1}^{T-1} \dfrac{1}{2}c_4^j\dfrac{dv_j}{d\gamma} + \dfrac{1}{2}2c_2^j\Big(\Delta x_j - c_1^j - c_2^j\gamma\Big)c_4^jv_j^2 - \dfrac{1}{2}\Big(\Delta x_j - c_1^j - c_2^j\gamma\Big)^2c_4^j2v_j\dfrac{dv_j}{d\gamma}\\
    &+\sum_{j=1}^{T-1}c_2^j\Big(\Delta x_j - c_1^j - c_2^j\gamma\Big)\dfrac{dv_j}{d\gamma}\\
    =& \sum_{j=1}^{T-1}\dfrac{1}{2}(c_4^j)^2v_j^2 + 2c_2^jc_4^j\Big(\Delta x_j - c_1^j - c_2^j\gamma\Big)v_j^2 - \Big(\Delta x_j - c_1^j - c_2^j\gamma\Big)^2(c_4^j)^2v_j^3\\
    =& \sum_{j=1}^{T-1} c_4^jv_j^2 \Bigg(\dfrac{1}{2}c_4^j + 2c_2^j\Big(\Delta x_j - c_1^j - c_2^j\gamma\Big) - \Big(\Delta x_j - c_1^j - c_2^j\gamma\Big)^2c_4^jv_j \Bigg).
\end{align}
Figure \ref{fig:likelihood-2ndderivative-distributions} shows the histogram of  $\dfrac{d^2f}{d\gamma^2}$ evaluated at the numerical root of $\dfrac{df}{d\gamma}$ on $[0,1]$.  
The second derivative is negative among for each of 100 instances of solving the optimization problem \eqref{eq:max-over-gamma}. 
We observe that the second derivatives are negative for all of the cases, which implies that the numerical root is reasonably presumed to be a maximum.
\comment{\begin{figure}[H]
  \centering
  \includegraphics[scale=0.19]{figures/likelihood_gamma_0_2ndderivative.png}
  \includegraphics[scale=0.19]{figures/likelihood_gamma_05_2ndderivative.png}
  \includegraphics[scale=0.19]{figures/likelihood_gamma_1_2ndderivative.png}
  \caption{Numerical distributions of the second derivative of the log-likelihood function $f(\gamma)$ evaluated at the numerical root on $[0,1]$ of the first derivative of $f(\gamma)$. \textbf{(A), (B), (C)} correspond to three different scenarios of density dependence $\gamma = 0$, $\gamma = 0.5$, $\gamma = 1$ respectively. The distributions are obtained from maximizing the log-likelihood function $f(\gamma)$ 100 times for each of the three $\gamma$ scenarios. We observe that the second derivatives are negative for all of the cases.}
  \label{fig:likelihood-2ndderivative-distributions}
\end{figure}}
\begin{figure}[H]
  \centering
  \includegraphics[scale=1]{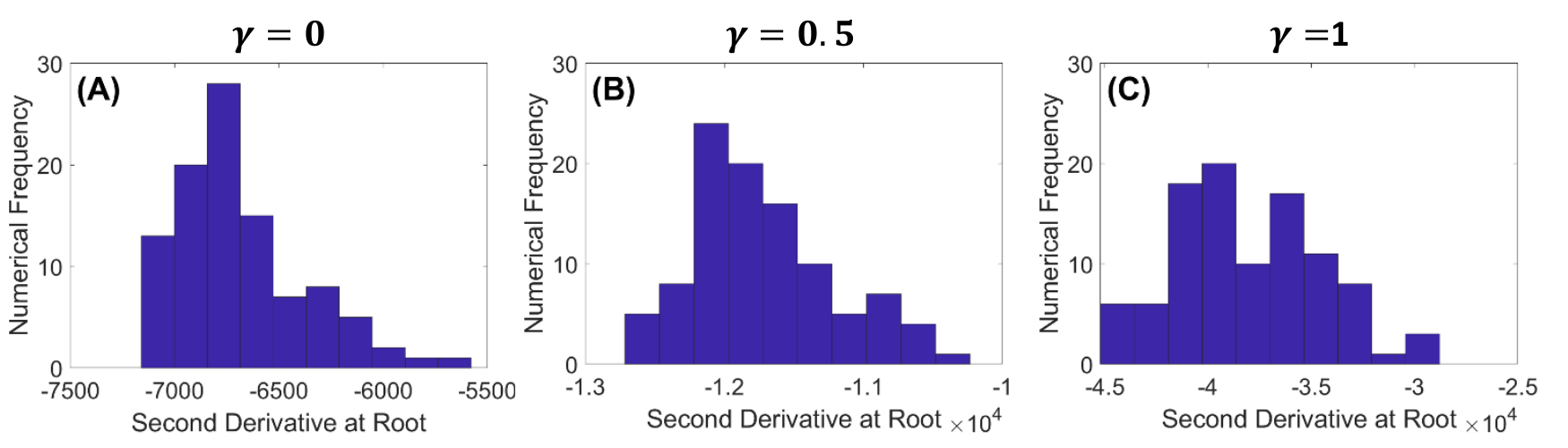}
  \caption{\textbf{The second derivative of the log likelihood is confirmed empirically to be negative at the numerical solution of the optimization problem.} We plot empirical distributions of the second derivative of the log-likelihood function $f(\gamma)$ evaluated at the numerical root on $[0,1]$ of the first derivative of $f(\gamma)$ for 100 cell number trajectories. \textbf{(A), (B), (C)} correspond to three different scenarios of density dependence $\gamma = 0$, $\gamma = 0.5$, $\gamma = 1$ respectively. The distributions are obtained from maximizing the log-likelihood function $f(\gamma)$ 100 times for each of the three $\gamma$ scenarios. We observe that the second derivatives are negative for all of the cases.}
  \label{fig:likelihood-2ndderivative-distributions}
\end{figure}
\noindent
We explicitly calculate the first derivative of $f(\gamma)$ below to find critical points:
\begin{align}
    \dfrac{df}{d\gamma} =& \sum_{j = 1}^{T-1}-\dfrac{1}{2}\dfrac{1}{x_j(b_{x_j}+d_{x_j})}x_j\dfrac{d}{d\gamma}(b_{x_j} + d_{x_j})\\
    &- \sum_{j = 1}^{T-1}\dfrac{1}{2}2\Big(\Delta x_j - x_j(b_{x_j} - d_{x_j})\Delta t \Big)\Delta t(-x_j)\dfrac{d}{d\gamma}(b_{x_j} - d_{x_j})\dfrac{1}{x_j(b_{x_j} + d_{x_j})}\dfrac{1}{\Delta t}\\
    &-\sum_{j = 1}^{T-1}\dfrac{1}{2}(\Delta x_j - x_j(b_{X_j} - d_{x_j})\Delta t)^2 \dfrac{-1}{\Delta t x_j^2(b_{x_j} + d_{x_j})^2}x_j\dfrac{d}{d\gamma}(b_{x_j} + d_{x_j})\\
    =& \sum_{j = 1}^{T-1}-\dfrac{1}{2}\dfrac{1}{(b_{x_j}+d_{x_j})}\dfrac{d}{d\gamma}(b_{x_j} + d_{x_j})\\
    &+ \sum_{j = 1}^{T-1}\Big(\Delta x_j - x_j(b_{x_j} - d_{x_j})\Delta t \Big)\dfrac{1}{(b_{x_j} + d_{x_j})}\dfrac{d}{d\gamma}(b_{x_j} - d_{x_j})\\
    &+ \sum_{j = 1}^{T-1}\dfrac{1}{2}(\Delta x_j - x_j(b_{X_j} - d_{x_j})\Delta t)^2 \dfrac{1}{\Delta t x_j(b_{x_j} + d_{x_j})^2}\dfrac{d}{d\gamma}(b_{x_j} + d_{x_j}),
\end{align}
where 
\begin{align}
    b_{x_j} + d_{x_j} &= \dfrac{b_0 - \gamma(r/K)x_j + \Big|b_0 - \gamma(r/K)x_j  \Big|}{2} + d_0 + (1-\gamma)(r/K)x_j\\
    \Rightarrow \dfrac{d}{d\gamma}\Big(b_{x_j} + d_{x_j}\Big)  &=  -\dfrac{(r/K)x_j}{2} - (r/K)x_j - \dfrac{1}{2}(r/K)x_j\dfrac{\Big|b_0 - \gamma(r/K)x_j  \Big|}{b_0 - \gamma(r/K)x_j}\\
    &= -\dfrac{3}{2}(r/K)x_j - \dfrac{1}{2}(r/K)x_j\dfrac{\Big|b_0 - \gamma(r/K)x_j  \Big|}{b_0 - \gamma(r/K)x_j},
\end{align}
and
\begin{align}
    b_{x_j} - d_{x_j} &= \dfrac{b_0 - \gamma(r/K)x_j + \Big|b_0 - \gamma(r/K)x_j  \Big|}{2} - d_0 - (1-\gamma)(r/K)x_j\\
    \Rightarrow \dfrac{d}{d\gamma}\Big(b_{x_j} - d_{x_j}\Big)  &=  -\dfrac{(r/K)x_j}{2} + (r/K)x_j - \dfrac{1}{2}(r/K)x_j\dfrac{\Big|b_0 - \gamma(r/K)x_j  \Big|}{b_0 - \gamma(r/K)x_j}\\
    &= \dfrac{1}{2}(r/K)x_j - \dfrac{1}{2}(r/K)x_j\dfrac{\Big|b_0 - \gamma(r/K)x_j  \Big|}{b_0 - \gamma(r/K)x_j}.
\end{align}
Using these expressions, we numerically obtain the root of the first derivative on the interval $[0,1]$.  

\end{document}